\documentclass[twocolumn,showpacs,aps,prb,a4paper]{revtex4-1}

\usepackage{amssymb,amsmath}
\usepackage{graphicx}

\usepackage[usenames,dvipsnames]{color}
\definecolor{MyOrange}{rgb}{1.0,0.5,0}
\definecolor{MyPurple}{rgb}{0.5,0,1}

\newcommand{\evib}{{e\text{-vib}}} 
\newcommand{\elvib}{electron-vibron\ } 

\newcommand{\GFs}{{\rm Green's functions}\ }

\newcommand{\SEs}{{\rm self-energies}\ }

\begin{document}

\title{Non-equilibrium inelastic electronic transport: Polarization
  effects and vertex corrections to the self-consistent Born
  approximation}

\author{L. K. Dash}
\email{louise.dash@york.ac.uk} 
\author{H. Ness}
\author{R. W. Godby} 
\affiliation{Department of Physics, University of York, York YO10 5DD, UK}
\affiliation{European Theoretical Spectroscopy Facility}
\date{\today}
 
\begin{abstract}
  We study the effect of electron-vibron interactions on the inelastic
  transport properties of single-molecule nanojunctions.  We use the
  non-equilibrium Green's functions technique and a model Hamiltonian
  to calculate the effects of second-order diagrams (double-exchange
  DX and dressed-phonon DPH diagrams) on the electron-vibration
  interaction and consider their effects across the full range of
  parameter space.  The DX diagram, corresponding to a vertex
  correction, introduces an effective dynamical renormalization of the
  electron-vibron coupling in both the purely inelastic and the
  inelastic-resonant features of the IETS. The purely inelastic
  features correspond to an applied bias around the energy of a
  vibron, while the inelastic-resonant features correspond to peaks
  (resonance) in the conductance.  The DPH diagram affects only the
  inelastic resonant features.  We also discuss the circumstances in
  which the second-order diagrams may be approximated in the study of
  more complex model systems.
\end{abstract}


\maketitle

\section{Introduction}
\label{sec:Introduction}

Junctions consisting of a single organic molecule between
two metallic leads hold great promise for future nanoscale devices,
where their potential applications include switches, transistors, and sensors.
Experimentally, it has proved difficult to control their production in
an atomistic manner, and so theoretical studies are crucial for a full
understanding of their behaviour.  It is known that inelastic effects
play an important role in the behaviour of such devices
\cite{Hipps:1993,Liu:2004,Kushmerick:2004,Yu:2004,Yu:2006,Chae:2006,
Beebe_JM:2007,Okabayashi:2008,
Gawronski:2009,Kim:2010}, but as yet we
lack a full understanding of the processes at play that will lead to a
complete interpretation of experimental results.

In this paper, we use a model-system nanojunction including electron-vibration
coupling
\cite{Hyldgaard:1994,Ness:1999, Ness:2001,
  Ness:2002a, Flensberg:2003, Mii:2003, Montgomery:2003b, Troisi:2003,
  Chen:2004, Lorente:2000, Frederiksen:2004, Galperin:2004,
  Galperin:2004b, Mitra:2004, Pecchia:2004, Pecchia:2004b,
  Chen_Z:2005, Paulsson:2005, Ryndyk:2005, Sergueev:2005, Viljas:2005,
  Yamamoto:2005, Cresti:2006, Kula:2006, Paulsson:2006, Ryndyk:2006,
  Troisi:2006b, Vega:2006, Caspary:2007, Frederiksen:2007,
  Galperin:2007, Ryndyk:2007, Schmidt:2007, Troisi:2007, Asai:2008,
  Benesch:2008, Paulsson:2008, Egger:2008, Monturet:2008,
  McEniry:2008, Ryndyk:2008, Schmidt:2008,
  Tsukada:2009,Loos:2009,Avriller:2009,Haupt:2009,
  Dash:2010,Ness:2010,Wang_H:2011,Garcia-Lekue:2011,Ueda:2011}
in order to
determine, by investigating the whole parameter space, what level of
diagrammatic expansion is appropriate to describe the
electron-vibration interaction in such junctions.  We cover the entire
parameter space, thus accounting for all the physical analogues to our
model.  The parameters we can vary correspond to the
lead-molecule-lead coupling, the electron-vibron
coupling strength, and the resonance of the electronic
level with the leads' electronic states

In general, an organic molecule-based nanojunction is unlikely to have
its HOMO or LUMO levels in alignment with the equilibrium
Fermi level of the leads, and so such a nanojunction will be dominated
by what we term the off-resonant regime with strong tunneling at low bias.  
Moreover, coupling between the leads and the central molecule is likely to 
be relatively weak, in the sense that the corresponding coupling to the leads
is much smaller than the corresponding hopping integrals in the
leads themselves.  
Conversely, a system consisting of a nanoconstriction in a gold wire 
will have
electronic levels in the constriction that are close to those of the
leads, and a larger coupling between the central region and the
leads, closer to the tight-binding hopping parameter of the leads.

This work will distinguish between these regimes and discuss for which
physical systems diagrams beyond the self-consistent Born
approximation (SCBA) become relevant for electron-phonon coupling.  We
study this using a full non-equilibrium Green's-function (NEGF)
technique \cite{Hyldgaard:1994,Mii:2003,Frederiksen:2004,
  Galperin:2004b, Mitra:2004, Pecchia:2004b, Chen_Z:2005, Ryndyk:2005,
  Sergueev:2005, Viljas:2005, Yamamoto:2005, Cresti:2006, Vega:2006,
  Egger:2008,Dash:2010} which allows us to study all the different
transport regimes in the presence of \elvib interaction.  Following
the spirit of many-body perturbation theory and Feynman diagrammatics,
we include the commonly-used SCBA diagrams as well as second-order 
diagrams in terms of the \elvib interaction.  A
detailed description of the formalism and the numerical implementation
of the NEGF code we have developed is given in
Ref.~[\onlinecite{Dash:2010}].  In this work we studied the
equilibrium and non-equilibrium electronic structures of the
nanojunctions in the presence of \elvib coupling.  In the present paper,
we now give and analyse results for the full non-equilibrium transport
properties, namely the non-linear $I(V)$ characteristics, the conductance
$G(V)=dI/dV$, and especially the IETS signal $d^2I/dV^2$ calculated
with our NEGF code.

The paper is structured as follows.  In Section \ref{sec:Theory} we
summarize the key aspect of our methodology detailed in
Ref.[\onlinecite{Dash:2010}].  Results for the effects of the second
order diagrams on the non-equilibrium non-linear transport properties
are presented in Section \ref{sec:Results}. They are separated into
the features we observe for the purely inelastic effects in the IETS
signal at bias equal to an integer multiple of the vibron energy, and
for the inelastic resonant features related to the vibron replica of
the electronic resonances.  Our conclusions are given in Section
\ref{sec:Conclusions}.  In addition, we explain in detail in Appendix
\ref{sec:Appendix-vertex} how one of the second-order diagrams
acts as a vertex correction to the Fock-like \elvib
diagram.

\section{Model and Theory}
\label{sec:Theory}

A fully atomistic description of the non-equilibrium inelastic
transport properties we wish to study is, unfortunately, beyond the
reach of current \emph{ab initio} methods.  Instead we use a model
system which retains the essential physics of the junction while
reducing the calculations to a tractable size.  A full description of
both the model and our methodology is given in
Ref.~[\onlinecite{Dash:2010}], and so we review only the most salient
features here.

We use the single-site single-mode model (SSSM),in which the central
molecule of the junction is represented by a single molecular level
coupled to a single vibrational mode.  We have already used this model
and discussed its validity in our previous study on the equilibrium
and non-equilibrium electronic structures of such a system coupled to
two electron reservoirs \cite{Dash:2010}.

The total Hamiltonian for the nanojunction is given by
\begin{equation}
  \label{eq:Htotal}
  H_{\text{total}} = H_L + H_R + V_{LC} + V_{CR} +
  H_C^e + H_{\text{vib}} + H_{\evib}.
\end{equation}
In this work, we represent the Hamiltonian of the left ($L$) and right
($R$) leads $H_{L,R}$ with a non-interacting tight-binding model with
semi-elliptic bands, although in principle it can take any valid form.
The hopping between leads and the central region is given by
$V_{\alpha C} = t_{0\alpha} \left( c^\dagger_\alpha d + d^\dagger c_\alpha \right)$, 
where $t_{0L,R}$ is the hopping integral between the
$\alpha=L,R$ lead and the central region.  The central region contains 
the electron-vibron interactions. We choose
that an electron couples linearly, via its density, to the
displacement of a single vibration mode.  The Hamiltonian for the
central region in the SSSM model is then given by
\begin{equation}
\label{eq:H_central}
\begin{split}
  H_C 
  & = H_C^e + H_{\text{vib}} + H_{\evib} \\
  & = \varepsilon_0 d^\dagger d + \hbar \omega_0 a^\dagger a +
  \gamma_0 (a^\dagger + a) d^\dagger d,
\end{split}
\end{equation}
where $d^\dagger$ ($d$) creates (annihilates) an
electron in the molecular level $\varepsilon_0$, which
is coupled to  the vibration mode 
of energy $\omega_0$ via the coupling constant $\gamma_0$.

A detailed analysis of the formalism of the non-equilibrium transport
properties from NEGF and for interacting system is provided
in Ref.~[\onlinecite{Ness:2010}]. 
The current $I_\alpha$ passing through each lead $\alpha$ is expressed in
terms of two time Green's functions\cite{Meir:1992}. It is transformed into frequency
representation for the steady-state regime to give:
\begin{equation}
  \label{eq:Current-Green's}
  I_\alpha = \frac{2e}{\hbar} \int \frac{{\rm d}\omega}{2\pi}\ 
  {\rm Tr}\{ 
  \Sigma_\alpha^<(\omega)\ G^>(\omega)-\Sigma_\alpha^>(\omega)\
  G^<(\omega)\}.
\end{equation}
We vary the applied bias by moving the chemical potentials of the left
and right leads.  With the equilibrium Fermi energies $\mu_L^{\text{
  eq}} = \mu_R^{\text{eq}} = \mu^{\text{eq}} = 0$, we introduce a
quantity $\eta_V$ such that $\mu_L = \mu^{\text{eq}} + \eta_V e V$ and
$\mu_R = \mu^{\text{eq}} - (1 - \eta_V) e V$ 
following Ref.~[\onlinecite{Datta:1997}].  In this way we can
create several forms for the potential drop across the junction.  By
setting $\eta_V = 1$, for example, we create an asymmetric drop whereby
$\mu_R$ remains constant while $\mu_L$ is changed, whereas $\eta_V =
0.5$ gives a symmetric potential drop where $\mu_L$ rises (lowers) as
$\mu_R$ lowers (rises) by the same amount.

It now remains to construct our non-equilibrium Green's functions
(details given in Ref.~\onlinecite{Dash:2010}).
The retarded and advanced Green's functions are calculated using a
Dyson equation
\begin{equation}
  \label{eq:Greens-Dyson}
  G^{r,a} = g^{r,a}_C + g^{r,a}_C\Sigma^{r,a}G^{r,a},
\end{equation}
while the greater ($G^>$) and lesser ($G^<$) Green's functions are obtained
from a quantum kinetic equation with the form
\begin{equation}
  \label{eq:Greens-greater-lesser}
  G^{>,<} = (1 + G^r \Sigma^r) g^{>,<}_C (1 + \Sigma^a G^a) + G^r
  \Sigma^{>,<} G^a,
\end{equation}
where $g_C$ is the \emph{non-interacting} Green's function of the isolated
central region.

\begin{figure}
  \centering
  \includegraphics[width=0.3\columnwidth]{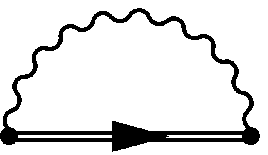}
  \includegraphics[width=0.2\columnwidth]{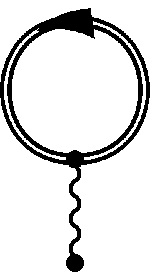}
  \caption{The Fock-(F)  and Hartree-like (H) diagrams}
  \label{fig:FockHartree}
\end{figure}

\begin{figure}
  \centering
  \includegraphics[width=0.3\columnwidth]{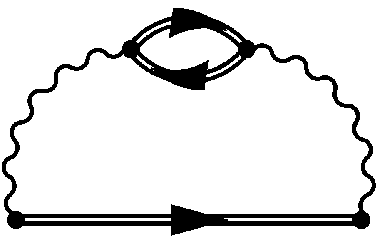}
  \includegraphics[width=0.3\columnwidth]{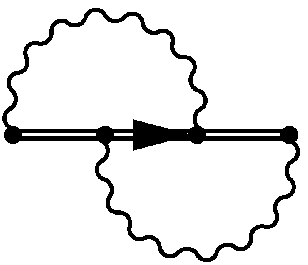}
  \caption{The dressed-phonon (DPH) and double exchange (DX) diagrams}
  \label{fig:DPH-DX}
\end{figure}

In this work we consider both first- and second-order contributions to
the electron-vibron coupling.  The first-order diagrams are shown in
Fig.~\ref{fig:FockHartree}, and, if calculated self-consistently,
equate to the commonly-used self-consistent Born approximation (SCBA).
We also make use of the two second-order diagrams, those which involve
two phonon excitations (Fig.~\ref{fig:DPH-DX}).  The first of these is
similar in structure to the $GW$ skeleton for electron-electron
interactions, and consists of a Fock-like diagram where the phonon is
dressed by a single electron-hole bubble (hence the appellation DPH
for dressed-phonon diagram, Fig.~\ref{fig:DPH-DX} left).  The second,
which we call the double-exchange (DX) diagram (Fig.~\ref{fig:DPH-DX}
right), includes two phonons simultaneously, with the second being
emitted before the first is reabsorbed. The DX diagram is part of the
skeleton diagrams corresponding to vertex corrections.  We use these
diagrams to construct expressions for the electron-vibron
self-energy \cite{Dash:2010} as the (total or partial) sum of each 
diagram: $\Sigma_\evib^{\text{total}} = \Sigma_\evib^{\text{F}} +
\Sigma_\evib^{\text{H}} + \Sigma_\evib^{\text{DPH}} +
\Sigma_\evib^{\text{DX}}$ .


Note that in order to handle numerically sharply peaked functions or strongly
discontinuous functions, we have found it necessary to include a very
small but finite imaginary part in our expression for the bare vibron
Green's function \cite{Dash:2010}.  This also allows us to perform
calculations with a smaller number of $\omega$-grid points, as long as
our imaginary part $\eta$ in the bare vibron Green's function is around two
to three times the $\omega$-grid spacing. We have already discussed in
detail the effects of the corresponding extra broadening on the lineshape of the
spectral functions and on the values of the linear conductance in
Ref.~[\onlinecite{Dash:2010}].

\section{Results}
\label{sec:Results}

In this section we present the effects of the second-order diagrams on
the full non-equilibrium transport properties of the nanojunction in
the presence of \elvib coupling.  Calculations of the \GFs are
performed with different levels of approximation for the \elvib
self-energies (Figs.~\ref{fig:FockHartree} and \ref{fig:DPH-DX}).
Fully self-consistent calculations using the first-order diagrams are
annotated SCBA, those using one or both second-order diagrams are
annotated SC(BA+DX), SC(BA+DPH) or SC(BA+DX+DPH) as appropriate. In
addition we have performed non-self-consistent second-order
corrections, i.e.  by using the SCBA \GFs to calculate the
second-order diagrams, we then determine the new \GFs without full
self-consistency. These calculations are annotated SCBA+DX or
SCBA+DPH.

The inelastic properties of the system are present in the current
$I(V)$ but are better represented by the second derivative of the
current $d^2I/dV^2$ as it is the signal that is directly measured
experimentally in the form of the inelastic electron tunneling
spectrum (IETS)\cite{Hipps:1993}.

The IETS curves present features, peaks or dips
\cite{Galperin:2004b,Ness:2010} at biases corresponding to the energy
of a specific excitation, in our case to the energy of one or several
excitations of the vibration mode. 
The peak feature is commonly associated with the opening of a new 
inelastic channel for the conductance of nanojunctions in the off-resonant
regime, i.e.~when
the electronic level $\varepsilon_0$ is sufficiently far from the
equilibrium Fermi level.
In the case of the resonant transport regime (when $\varepsilon_0$
is close to $\mu^{\text{eq}}$), a dip feature is obtained in the IETS.
It is associated with \elvib backscattering effects and a decrease in 
the conductance at the threshold bias.
Furthermore, being the derivative of the
conductance, the IETS curves also present features at biases
corresponding to peaks in the conductance.  We have found
\cite{Dash:2010} that in order to get a better aspect ratio for the
IETS features corresponding to vibron excitations, it is more
convenient to normalize the IETS curves by the dynamical conductance, i.e.~$[d^2I/dV^2]/[dI/dV]=d/dV \ln G(V)$
as in  Refs.~\onlinecite{Hipps:1993,Kushmerick:2004,Yu:2004,Yu:2006,Beebe_JM:2007,Okabayashi:2008,Kim:2010}.

We divide our results into two sections; the first for purely
inelastic features, and the second for inelastic features associated
with the electron resonance effects.  The first category corresponds
to features observed in the IETS signal at bias equal to an integer
multiple of the vibron energy $n\omega_0$, i.e. a tunneling electron
excites $n$ vibrons.  The second category corresponds to inelastic
resonant tunneling via the vibron replica associated with the main
electronic resonance at $\tilde{\varepsilon}_0 \sim \varepsilon_0 -
\gamma_0^2/\omega_0$, and hence are observed in the IETS signal for
biases $V \sim \tilde{\varepsilon}_0 \pm n\omega_0$.  We will see
below that the second-order diagrams have different effects on the
IETS features depending upon their correspondence to one of these two
categories.

\subsection{Purely inelastic features}
\label{sec:Pure-inelastic-features}

We consider first of all the off-resonant regime.  In this limit
the IETS features associated with inelastic resonant tunneling
are sufficiently far (or sufficiently small for the higher vibron
replica) from the inelastic feature at $V=\omega_0$. We can thus
avoid a superposition of the two different kinds of features.

\begin{figure}
  \centering
  \includegraphics[width=\columnwidth]{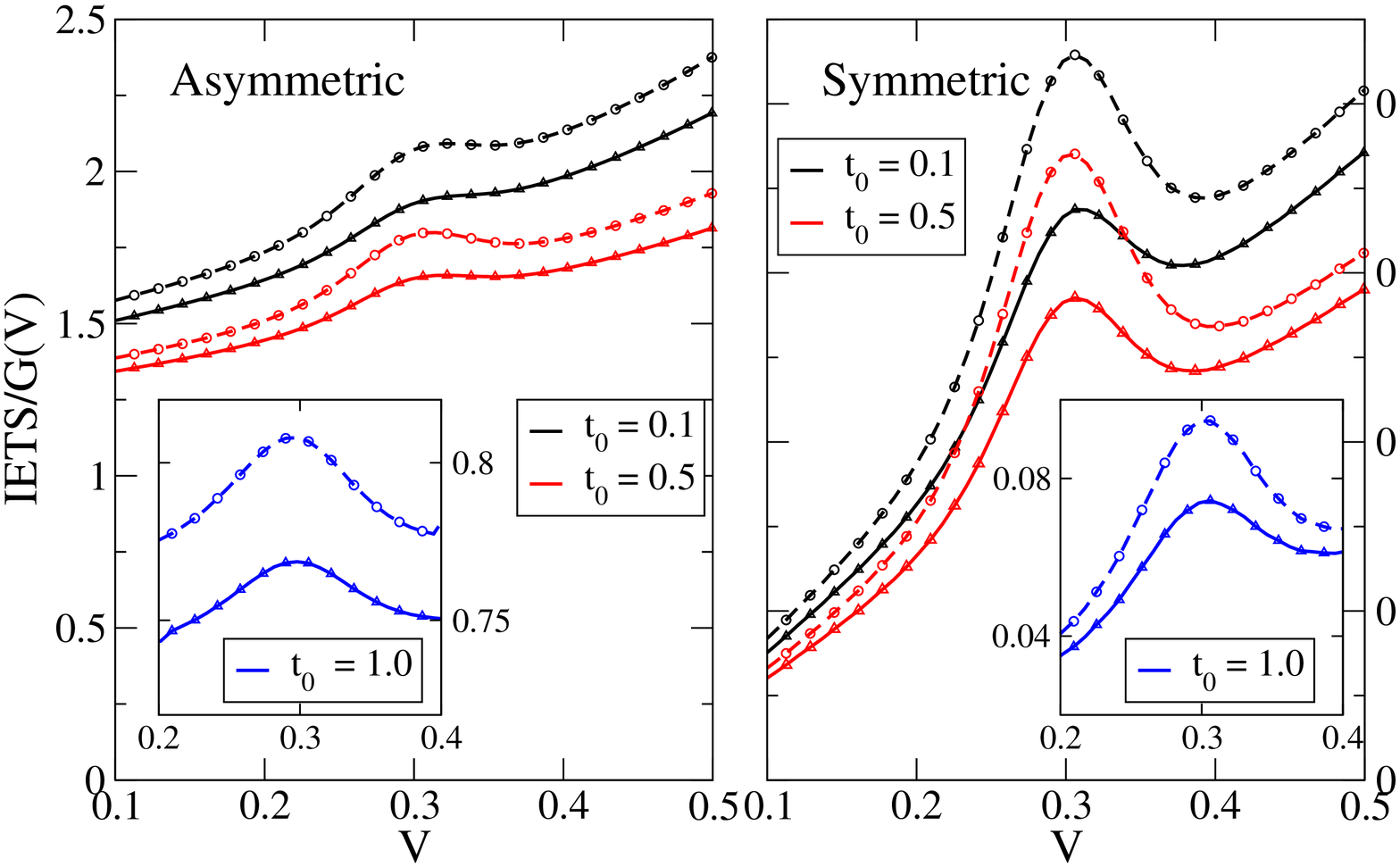}
  \caption{(Color online) The inelastic peak in the IETS around
    $V=\omega_0=0.3$ for an asymmetric potential drop ($\eta_V =
    1.0$, left panel) and a symmetric potential drop ($\eta_V = 0.5$,
    right panel). Results are shown for three different lead-molecule
    coupling
    parameters $t_0=t_{0L,R}$ (black, red, and blue lines), 
    and for two different \elvib
    couplings (medium $\gamma_0 = 0.195$, solid lines and triangles;
    strong $\gamma_0 = 0.25$, dashed lines and circles).  Lines
    represent calculations at the SCBA level, symbols SC(BA+DPH), from
    which we can see there is no distinguishable contribution to the
    inelastic peak from DPH for any parameter set. The other parameters
    are $\varepsilon_0=+1.5, \omega_0=0.3, \eta=0.030$.}
  \label{fig:w0peak}
\end{figure}

Results are presented in Fig.~\ref{fig:w0peak} for different
electron-vibron couplings and for both symmetric and asymmetric
potential drops.  We note that, as would be expected intuitively, the
size of the feature increases with the electron-phonon coupling
$\gamma_0$---in fact the height with respect to the baseline is
proportional to $\gamma_0^2$.  As $t_{0L,R}$ is increased, the IETS
signal decreases in overall magnitude, although the feature at
$\omega_0$ itself remains clear (Fig.~\ref{fig:w0peak} inset).

In the off-resonant regime, we do not find any difference to the
curves when the second-order dressed-phonon (DPH) diagram is included.
Possible reasons for this are discussed in section
\ref{sec:Electron-resonance}.

\begin{figure}
  \centering
  \includegraphics[width=\columnwidth]{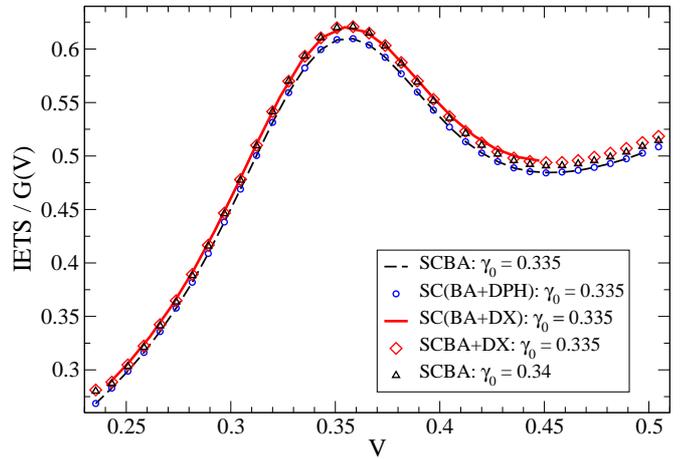}
  \caption{(Color online) Normalized IETS signal for the inelastic
    peak at $\omega_0 = 0.35$.  Results for SCBA and SC(BA+DPH) are
    virtually identical.  However, including the DX diagram, either
    self-consistently (solid red line) or non-self-consistently (red
    diamonds) raises the amplitude of the peak.  This effect can 
    be approximated by an SCBA calculation done with a larger $\gamma_0$
    (black triangles), indicating that DX generates an effective
    renormalization of $\gamma_0$. The other parameters are
    $\varepsilon_0=+1.5, \omega_0=0.35, t_{0L,R}=0.10, \eta=0.039,
    \eta_V=0.5$. }
  \label{fig:SCBA+DX}
\end{figure}
\begin{figure}
  \centering
  \includegraphics[width=\columnwidth]{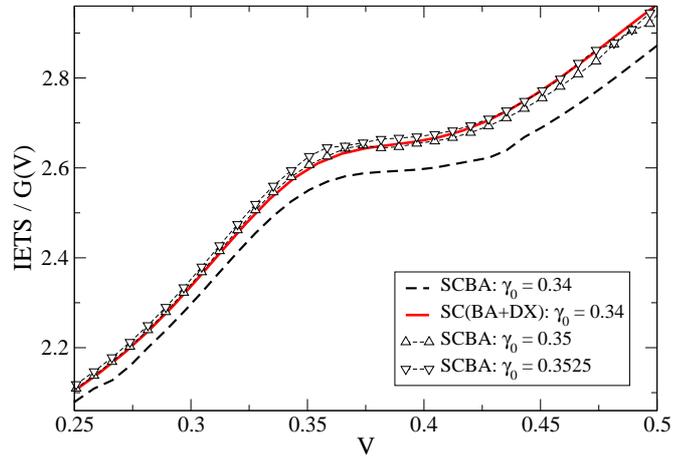}
  \caption{(Color online) Normalized IETS signal for biases around
    $\omega_0$ showing different SCBA fits for the renormalized
    electron-vibron coupling. The SCBA (dashed line) and SC(BA+DX)
    (solid red line) are shown for $\gamma_0 = 0.34$, together with
    SCBA calculations for $\gamma_0 = 0.35$ and $0.3525$.  Although
    these provide a good approximation, neither gives an exact fit
    to the lineshape of the SC(BA+DX) curve. The other parameters are
    $\varepsilon_0=+1.5, \gamma_0=0.34, \omega_0=0.35, t_{0L,R}=0.10,
    \eta=0.039, \eta_V=1.0$.}
  \label{fig:IETS_badx_andfit}
\end{figure}

We also note that the results are dependent on the symmetry of the
potential drops at the left and right contact.  As shown in the
separate panels of Fig.~\ref{fig:w0peak}, the curves for symmetric
potential drops have a much lower baseline than for the asymmetric
potential drops, although the inelastic features have the same
lineshape, position and magnitude (see Figs.~\ref{fig:w0peak},
\ref{fig:SCBA+DX} and \ref{fig:IETS_badx_andfit}).  This is because
(for positive bias) in the asymmetric case, $\mu_L$ is rising bias
while $\mu_R$ is kept fixed at its equilibrium value. Thus $\mu_L$ is
approaching the electron resonance twice as fast as in the symmetric
case, where $\mu_L$ rises at the same rate as $\mu_R$ falls.  The
symmetric case therefore contains much less effect from the tail of
the main electron resonance, and allows us to further isolate the
purely inelastic part of the IETS signal.


We now consider the effect of the double-exchange diagram
(Fig.~\ref{fig:DPH-DX} right) on the inelastic peak at $\omega_0$.
Calculation of $\Sigma_\evib^{\text{DX}}$ is extremely computationally
intensive, as it can be reduced neither to a simple convolution
product, nor even to a simple double-convolution product
\cite{Dash:2010}.  Calculations for the integration in the energy
representation of the \GFs and the \SEs scale as the cube of the number
$N_\omega$ of points in the energy-grid \cite{Dash:2010}.  Moreover,
as the finite imaginary part $\eta$ we have included in the vibron
Green's function causes extra broadening, if taken too large it washes
out much of the effect of the DX diagram.  In order to decrease $\eta$
we need to increase the number of grid points, and thus, even for our
minimal model, a fully self-consistent calculation of the
double-exchange diagram becomes intractable with $N_\omega > 2000$. 
Hence for the DX calculations, we usually work with $N_\omega \sim
1500 - 2000$, giving $\eta \sim 0.03 - 0.04$ for the total spectral
width considered in our calculations.

\begin{figure}
  \centering
  \begin{minipage}[t]{0.35\columnwidth} 
    \centering
      \includegraphics[width=\textwidth]{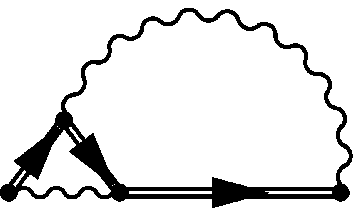}
      \vspace{0.1\textheight}
    \end{minipage}
   \hspace{0.08\textwidth}  
  \begin{minipage}[t]{0.35\columnwidth}
      \vspace{-0.073\textheight} \centering
      \includegraphics[width=\textwidth]{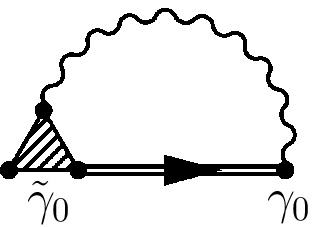}
    \end{minipage}
\vspace{-0.073\textheight}
  \caption{Second order double exchange (DX) diagram re-expression
    as a vertex correction to the Fock-like diagram, with a renormalized
    $\gamma_0 \rightarrow \tilde{\gamma}_0$ vertex.}
  \label{fig:DX-Feynman-mod}
\end{figure}

We work around this by calculating the effects of the DX diagram both
in a self-consistent manner SC(BA+DX) or non-self-consistenly as a
second-order correction to the SCBA result.

The results of this are plotted in Fig.~\ref{fig:SCBA+DX}.  We see
that the effect of the DX diagram is to both increase the height of
the feature and to raise its baseline.  For the set of parameters shown
the self-consistency in the DX diagram calculations is not crucial.

We see that the effect of DX can also be reproduced with a SCBA
calculation in which we increase the value of $\gamma_0$.  Therefore
the DX diagram has the effect of renormalizing $\gamma_0$ as it is
part of the skeleton family of vertex correction, as shown in
Fig.~\ref{fig:DX-Feynman-mod}.  We can thus approximate the effect of
the DX diagram in the IETS with a Fock-like diagram with one
renormalized vertex $\tilde{\gamma}_0$.

The amplitude of the peak
at $\omega_0$, instead of varying as $\gamma_0^2$, therefore now
depends on $\tilde{\gamma}_0\gamma_0$, and so we can define an
\emph{effective} electron-vibron coupling constant $\bar{\gamma}_0 =
\sqrt{\tilde{\gamma_0}\gamma_0}$, with $\bar{\gamma}_0 > \gamma_0$.
This allows us to make a more quantitative analysis of
$\tilde{\gamma_0}$ by fitting the SCBA+DX curve to an SCBA curve with
electron-vibron coupling $\bar{\gamma}_0$, as shown in
Fig.~\ref{fig:SCBA+DX}.  

\begin{figure}
  \centering
  \includegraphics[width=\columnwidth]{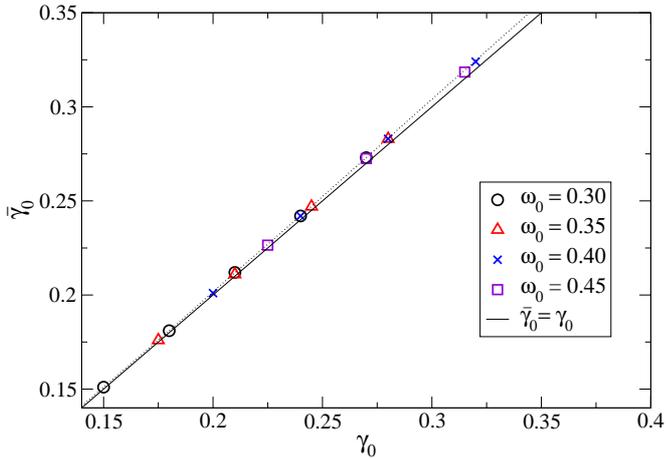}
  \caption{(Color online) Effective static renormalized
    electron-vibron coupling parameter $\bar{\gamma_0}$ as a function
    of the nominal coupling parameter $\gamma_0$, from a fit of
    several sets of data comparing SCBA+DX curves to SCBA (as in
    Fig.~\ref{fig:SCBA+DX}).  The data for all three values of
    $\omega_0$ lie on a straight line with slope 1.03 (the straight
    line with slope 1 is shown for comparison), implying that the
    vertex correction increases the effective \elvib coupling by 3\%.}
  \label{fig:gamma0correlation}
\end{figure}

Furthermore, we can study how $\tilde{\gamma}_0$ and $\gamma_0$ are
correlated by performing a series of calculations for different values
of the parameters $\gamma_0$ and $\omega_0$, and then fitting the SCBA+DX 
results to those from SCBA calculations.  The results of this are shown in
Fig.~\ref{fig:gamma0correlation}, from which we can see that, to a
first approximation, the DX diagram consistently raises the effective
electron-vibron interaction by approximately 3\% within the range of 
parameters we used. 

However, we would like to point out that although the apparent effects of
the DX diagram on the IETS signal is to renormalize the coupling
constant $\gamma_0$, the reality is much more subtle. In Appendix
\ref{sec:Appendix-vertex} we discuss in detail the renormalization
effect of the DX diagram in terms of vertex corrections, and we
show that such vertex corrections do not simply correspond to a
mapping of the SSSM Hamiltonian onto a similar Hamiltonian with a
static renormalization of the \elvib coupling constant, i.e.
$\gamma_0 \rightarrow \bar\gamma_0 = \gamma_0 +
\vert\Delta\vert$. Rather, the vertex correction actually generates a
dynamical renormalization of the \elvib coupling constant, i.e.
$\gamma_0 \rightarrow \bar\gamma_0(\omega,\omega')$.

This can be seen more clearly by considering SCBA calculations, for 
different values of the electron-vibron coupling parameter, 
and checking which of such calculations 
correspond the best to a SC(BA+DX) calculation. The result is shown in
Fig.~\ref{fig:IETS_badx_andfit}.  Although the difference between the
best SCBA fits and the SC(BA+DX) are not large, it is quite clear that
a renormalized SCBA calculation does not provide exactly the same
lineshape as a full SC(BA+DX) calculation for all the range of biases $V$ 
around the inelastic peak at $\omega_0$.

\begin{figure}
  \centering
  \vspace{2\baselineskip}
  \includegraphics[width=\columnwidth]{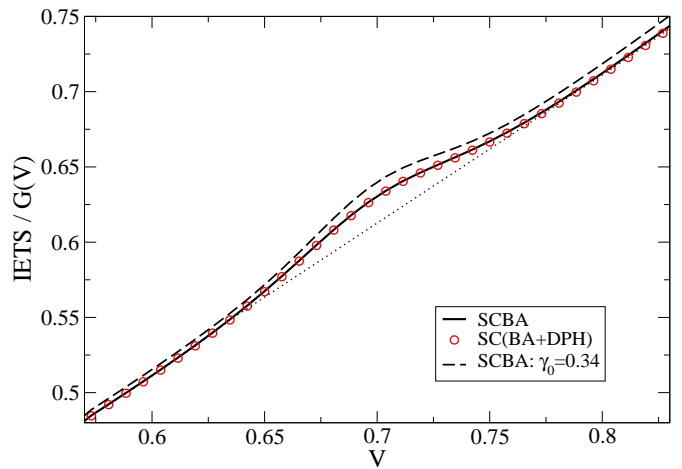}
  \caption{(Color online) Normalized IETS signal for biases around
    $2\omega_0$ for different levels of approximation: the SCBA (solid
    black line) and SC(BA+DPH) (red circles) are virtually identical,
    as for the feature at $\omega_0$. An SCBA calculation
    approximating the effect of DX is also included (dashed line). A
    straight baseline is included for clarity.  The other parameters
    are $\varepsilon_0=+1.5, \gamma_0=0.335, \omega_0=0.35,
    t_{0L,R}=0.10, \eta=0.009, \eta_V=0.5$.}
  \label{fig:offres_IETSsidebandpeak}
\end{figure}

Finally we expect to see a peak feature at $V \sim 2\omega_0$ in the IETS for
the off-resonant transport regime. This peak feature is the
two-vibration excitation equivalent of the feature observed at $V =
\omega_0$. Since this a higher-order process, the amplitude of the
feature should be $\gamma_0^2$ times smaller than the feature at $V
=\omega_0$. This feature has a rather small amplitude for all the \elvib coupling
constants we have considered in this work since $\gamma_0 < 1$.  An
example of a close-up of the IETS feature around $V \sim 2\omega_0$ is
given in Figure~\ref{fig:offres_IETSsidebandpeak}.  We find that the
amplitude of the peak with respect to the linear baseline is indeed
approximately $\gamma_0^2=0.112$ (one order of magnitude) smaller than
the corresponding amplitude of the peak at $V = \omega_0$ (shown in Fig.~\ref{fig:SCBA+DX}).
Once more we find that the effects of the DPH diagram are
negligible for this part of the IETS signal. In
Fig.~\ref{fig:offres_IETSsidebandpeak} we also include the results of
a SCBA calculation for a larger coupling $\gamma_0=0.34$ which mimics the 
renormalization effects of the DX diagram as discussion above.

\subsection{Inelastic resonant features}
\label{sec:Electron-resonance}

\begin{figure}
  \centering
 \includegraphics[width=\columnwidth]{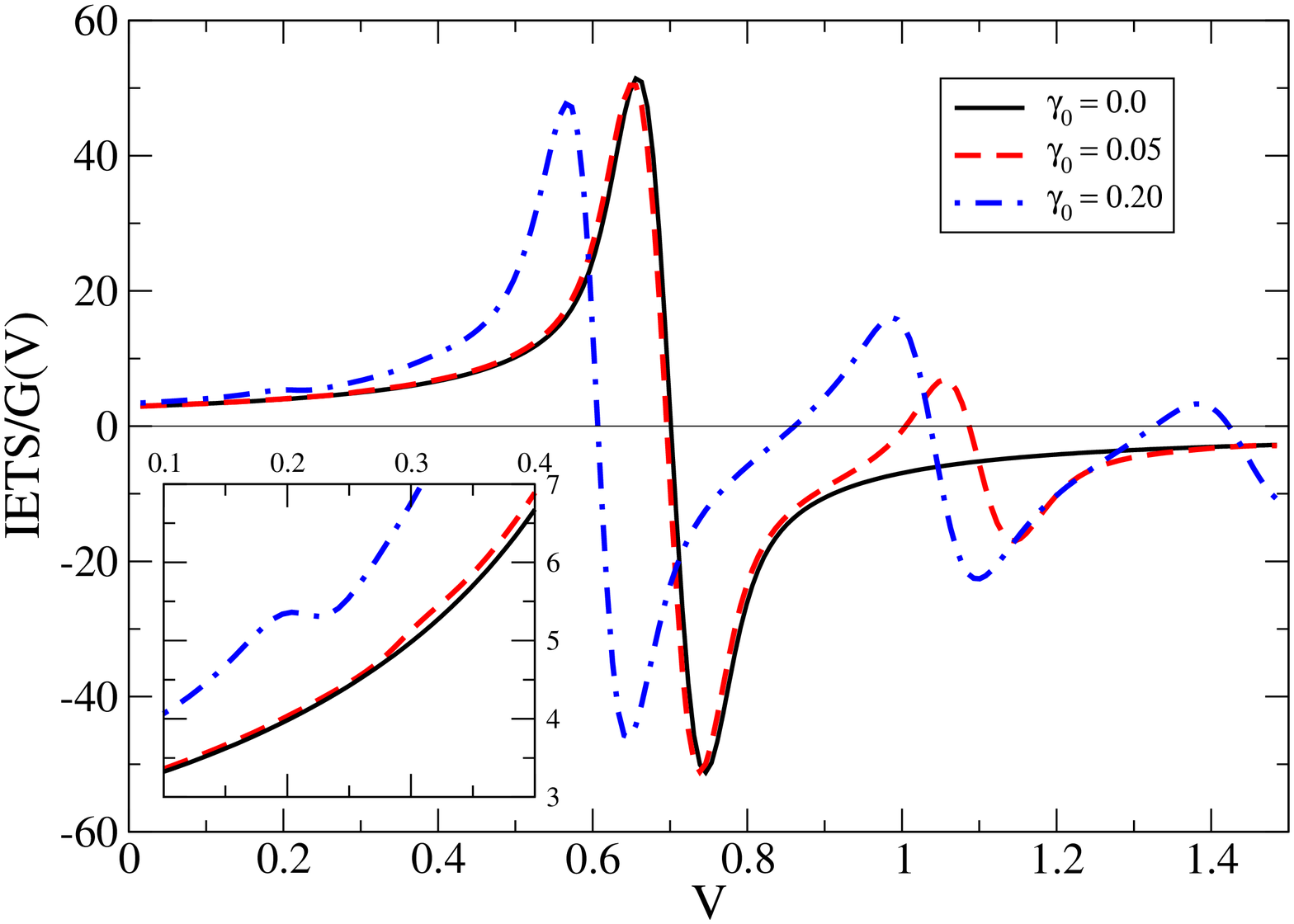}
 \\  \vspace{3\baselineskip}
   \includegraphics[width=\columnwidth]{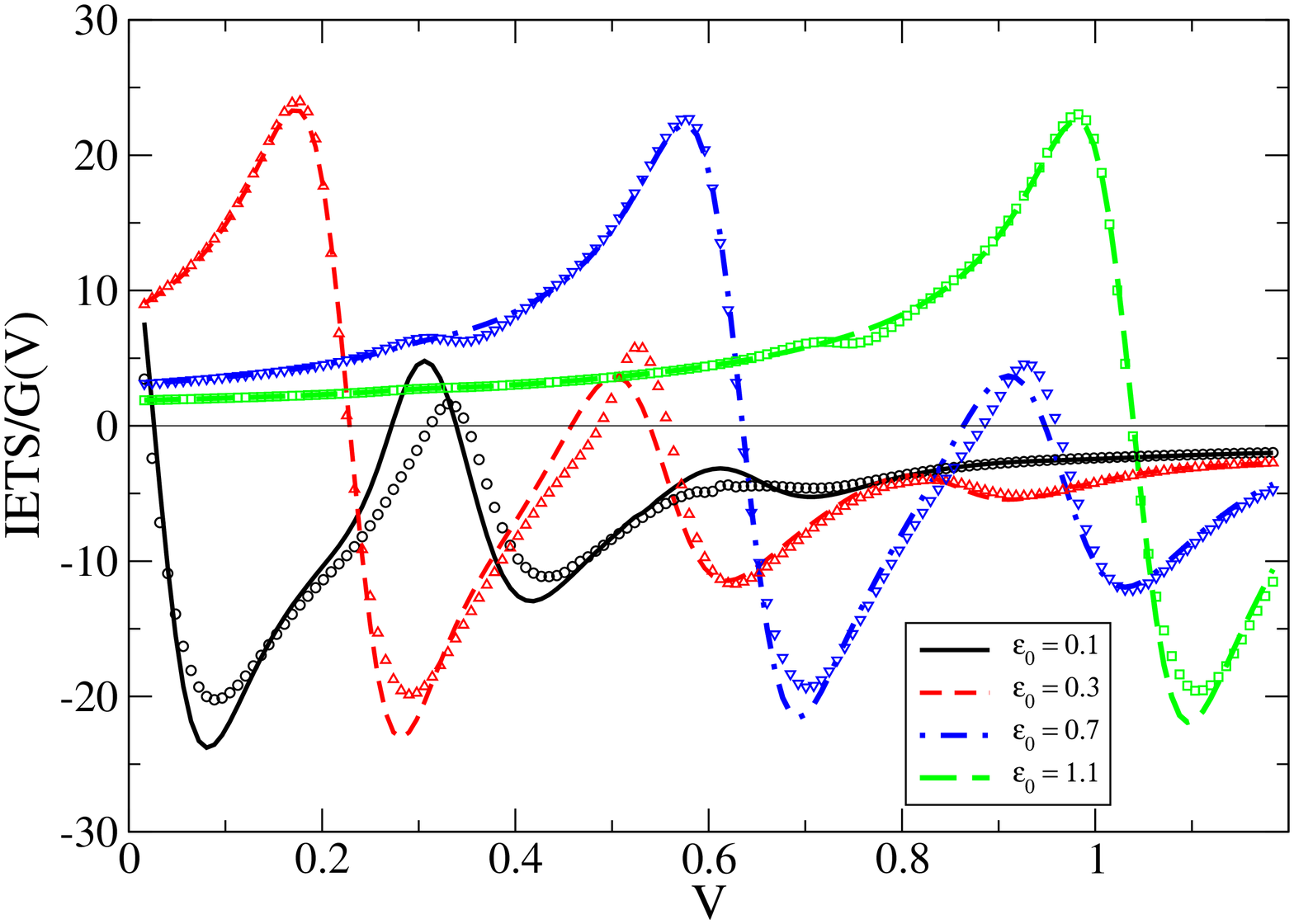}
   \caption{(Color online) Normalized IETS signal. Top: Inelastic
     resonant features for different values of $\gamma_0$.  With no
     electron-vibron coupling (solid black line), there is a single
     feature at $\varepsilon_0 = 0.7$.  With electron-vibron coupling
     (red and blue dashed lines), the main peak moves to
     $\tilde{\varepsilon}_0 \sim \varepsilon_0 - \gamma_0^2/\omega_0$
     and side-band peaks appear. The inset shows a tiny feature at
     $\tilde{\varepsilon}_0-\omega_0$ that only occurs for $t_0
     \lesssim 0.1$ in SCBA calculations but reappears for all $t_0$
     once DPH is included.  Bottom: The inelastic resonant features
     for various $\varepsilon_0$ at $\gamma_0 = 0.15$ for both SCBA
     (lines) and SC(BA+DPH) (symbols) calculations, showing the
     increasing influence of DPH in the resonant transport regime
     $\varepsilon_0\rightarrow 0$.  The other parameters are: $t_0 =
     0.2, \omega_0 = 0.3, \eta_V = 1.0, \eta = 0.03$. }
  \label{fig:resonance-peaks}
\end{figure}

The main electron-resonance peak occurs at the polaron-shifted value
of $\tilde{\varepsilon}_0 \sim \varepsilon_0 - \gamma_0^2/\omega_0$,
and consists of a peak-dip feature in the IETS, as it corresponds to a
resonant peak feature in the conductance.  
With no electron-phonon coupling ($\gamma_0 = 0$),
the IETS curve has no features other than the one corresponding to the
resonant transmission in the conductance at $V=\varepsilon_0$.  Once
the electron-phonon coupling is turned on, phonon side-band peaks
emerge in the spectral function at energies
$\omega=\tilde{\varepsilon}_0 \pm n\omega_0$.  The $+n\omega_0$
features correspond to phonon emission (vibration excitation) by an
electron, while the $-n\omega_0$ features correspond to
phonon-emission by a hole.  In the IETS, they appear as peak-dip
features (derivatives of an inelastic resonant peak in the
conductance) with amplitude decreasing as the bias is further
increased.  At lower biases however, there are no features at
$\omega=\tilde{\varepsilon}_0 - n\omega_0$ at the SCBA level except
for very low values of the coupling to the leads ($t_{0L,R} \lesssim
0.1$, Fig.~\ref{fig:resonance-peaks} top inset).

Including the DPH diagram, however,
introduces a small peak-dip features at $\tilde{\varepsilon}_0 -
\omega_0$ just below the main peak-dip feature at $V \sim \tilde{\varepsilon}_0$
in the IETS, as shown in Fig.~\ref{fig:resonance-peaks}).  
The DPH diagram also has a strong influence on the
lineshape of the other phonon side-band peaks above $\tilde{\varepsilon}_0$, 
increasingly so as
$\varepsilon_0$ is brought within range of the equilibrium chemical
potentials (i.e. $\tilde{\varepsilon}_0 \rightarrow 0$ in
Fig.~\ref{fig:resonance-peaks} bottom).

\begin{figure}
  \centering
  \includegraphics[width=\columnwidth]{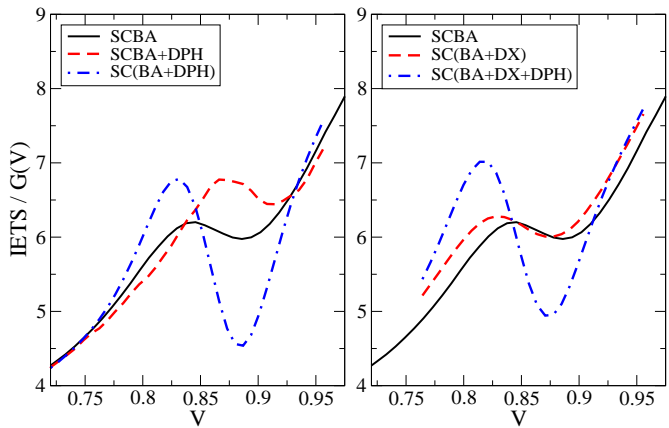}
  \caption{(Color online) Normalised IETS signal for biases around the
    first vibration side-band peak $V \sim \tilde{\varepsilon}_0
    -\omega_0 = 1.19 - 0.4 \sim 0.8$.  Calculations performed for
    different approximations for the electron-vibron self-energy:
    (left panel) SCBA (black solid line), SCBA plus second order DPH
    correction (dashed red line), SC(BA+DPH) (dot-dashed blue line);
    (right panel): SCBA (solid black line), SC(BA+DX) (dashed red
    line) and SC(BA+DX+DPH) (dot-dashed blue line).  The other
    parameters are $\varepsilon_0=+1.5, \gamma_0=0.35, \omega_0=0.4,
    t_{0L,R}=0.09, \eta=0.039, \eta_V=1$.}
  \label{fig:electronres_sidebandpeak}
\end{figure}

We now consider the combined effects of the DPH diagram as well as of
the DX diagram on the specific case of the IETS feature at
$\tilde{\varepsilon}_0 -\omega_0$. This is shown in Figure
\ref{fig:electronres_sidebandpeak}.  The DPH diagram increases strongly
the peak-dip feature (at $V \sim \tilde{\varepsilon}_0 -\omega_0$)
obtained from SCBA calculations at medium/strong electron-vibron
coupling.  Note here the importance of the self-consistency in the
calculations: the second-order DPH diagram calculated as a second-order
correction to SCBA (Fig.~\ref{fig:electronres_sidebandpeak} left) gives a
completely wrong feature in the IETS.

Interestingly, the self-consistent calculation with the DX diagram
seems to give a similar feature to that observed in the SCBA
calculations, but slightly shifted towards lower bias
(Fig.~\ref{fig:electronres_sidebandpeak} right). This is completely
consistent with the renormalization effects of the electron-vibron
coupling by the DX diagram as discussed in the previous
section. Indeed, the DX diagram renormalizes the coupling $\gamma_0$
towards a higher value $\bar\gamma_0$. Consequently the
renormalization of the molecular level by $- \bar\gamma_0^2/\omega_0$
is more important than for SCBA calculations, and thus the feature is
moved towards lower bias.

The calculations performed with both DX and DPH diagrams
(Fig.~\ref{fig:electronres_sidebandpeak} right) generate a hybrid
feature in the IETS in comparison to individual calculations with the
second-order diagrams.  However the new IETS is not simply obtained by
a linear superposition of the individual effects of the DPH and DX
diagrams.

It might at first seem strange that the DPH self-energy is negligble
at $V = n\omega_0$, where one might expect it to be influential, but that
it has a significant effect at biases $\gtrsim \tilde{\varepsilon}_0$.
While this is to some extent related to the strength of the electron-vibron
coupling, there is another, more important, underlying
cause. The DPH diagram involves an electron-hole bubble, and so for
this diagram to become relevant, simultaneous electron and hole states
must be available.  This is not the case when the spectral
functions of the coupled electron-vibron system are mostly empty or mostly
filled.  When the bias is significantly low and both Fermi levels
$\mu_{L,R}$ are below the electron resonance level, these excitations
are inaccessible and so there is no effects from the DPH diagram.  
Once the bias is increased to within range of $\tilde\varepsilon_0$, or
$\tilde\varepsilon_0 \pm n\omega_0$ however, these electron-hole states
become accessible and the DPH diagram becomes influential (unless the
spectral function is mostly filled).  This is also borne out by the
increasing contribution from DPH to the lineshape of the phonon
side-band peaks as the electron level $\varepsilon_0$is decreased, moving
DPH's sphere of influence to lower and lower biases (see lower panel
in Fig.~\ref{fig:resonance-peaks}).

\subsection{Summary over the entire parameter range}
\label{sec:Summary-parameter}

\begin{figure}
  \centering
  \begin{minipage}[t]{0.9\columnwidth} 
    \centering
      \includegraphics[width=0.49 \textwidth]{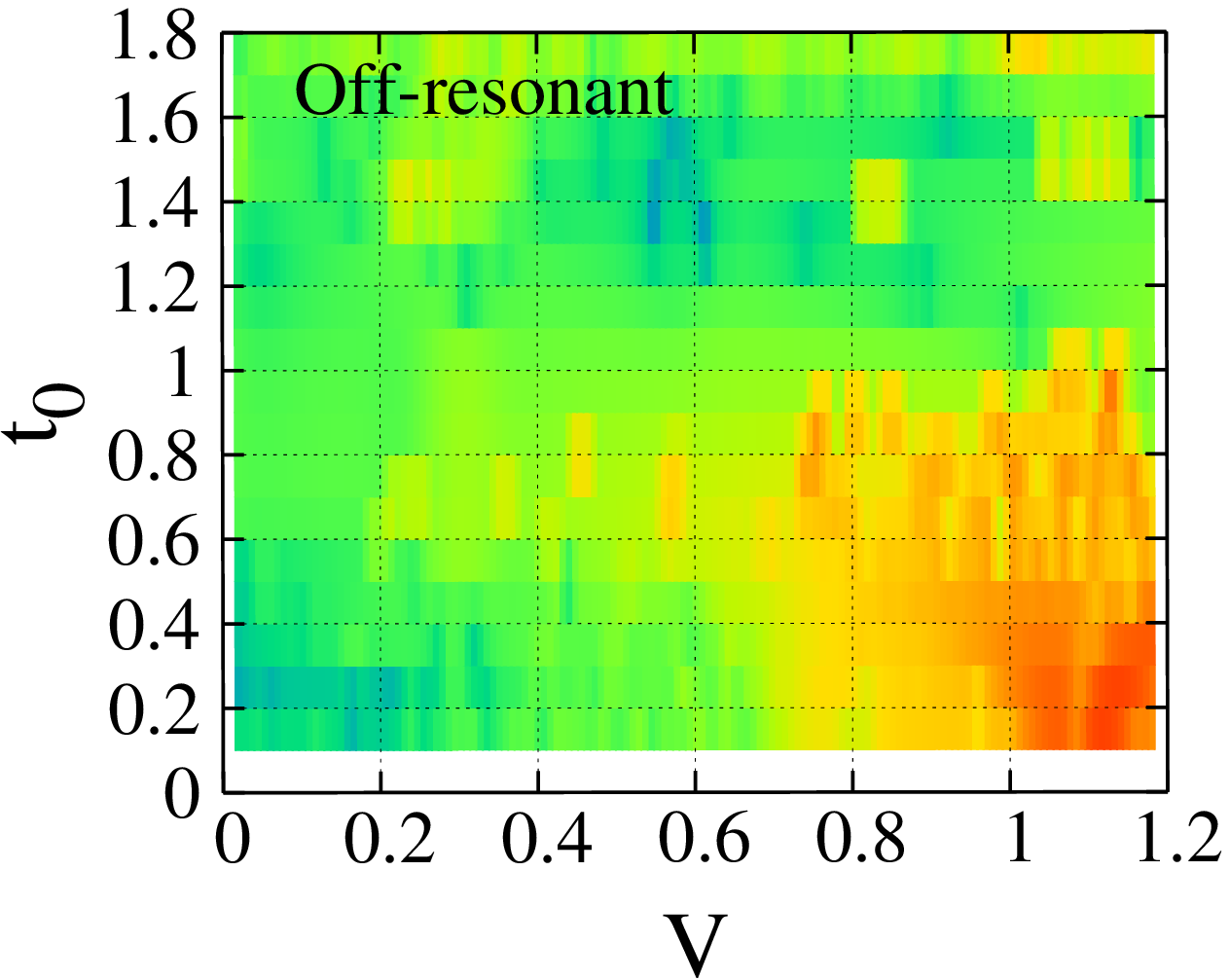}
      \includegraphics[width=0.49\textwidth]{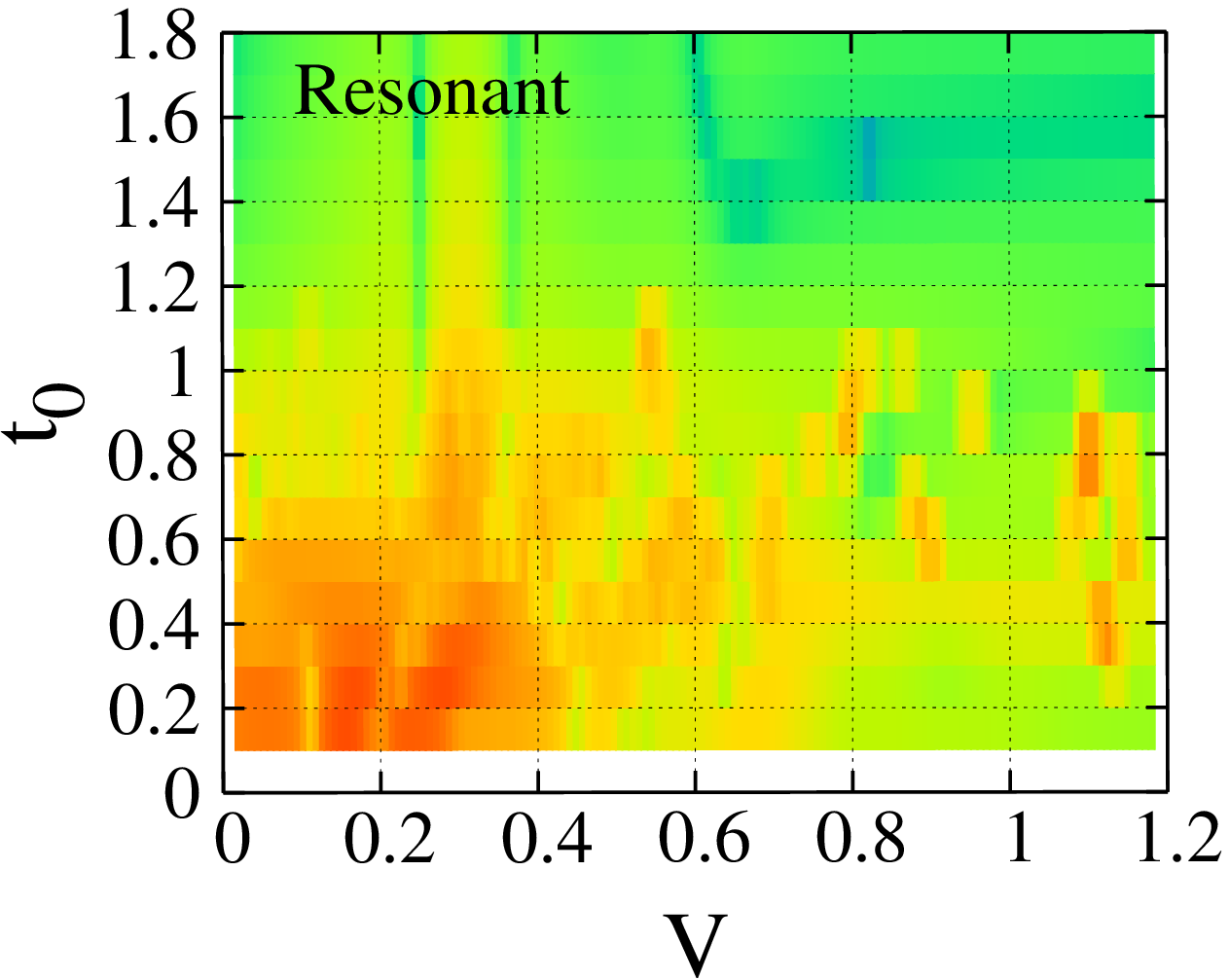}
      \vspace{0.11\textheight}
    \end{minipage}
  \begin{minipage}[t]{0.08\columnwidth}
      \vspace{-0.11\textheight} \centering
      \includegraphics[width=0.85\textwidth]{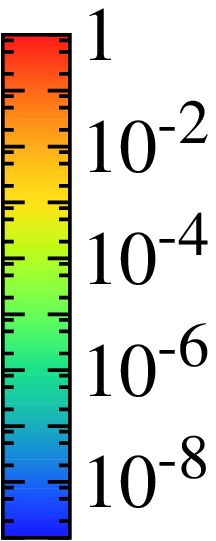}
    \end{minipage}
\vspace{-0.13\textheight}
\caption{(Color online) Maps generated from multiple calculations of
  SCBA and SC(DA+DPH) IET spectra for the off-resonant (left panel)
  and resonant (right panel) regimes showing the normalized absolute
  ratio of the SCBA and SC(BA+DPH) spectra $\left| 1 -
    \frac{\text{IETS}_{\text{SC(BA+DPH)}}}{\text{IETS}_{\text{SCBA}}}
  \right|$. It gives a near-zero result (blue and green areas) when
  the two calculations give the same spectrum, and a positive number
  (red areas, i.e. bottom right corner for SCBA and bottom left corner
  for SC(BA+DPH)) where there is a substantial difference between the
  two. The other parameters are $\varepsilon_0 = 1.5 / 0.0$
  (off-resonant/resonant), $\gamma_0 = 0.195, \omega_0 = 0.3$.}
  \label{fig:maps}
\end{figure}

Having examined the role of the second-order diagrams in detail
for characteristic selected sets of parameters, we
now show results across the entire parameter range.  In order to
present this in a concise manner, we have compiled maps of our IETS results
comparing SCBA calculations to those with SC(BA+DPH) indicating in
which regimes the DPH self-energy has a significant effect. For the
off-resonant regime (Fig.~\ref{fig:maps} left), we can see that the
greatest effect of DPH is apparent at higher bias (i.e. approaching
the electron resonance $\varepsilon_0$) and when the
lead-molecule-lead coupling is small.  For the resonant case, however,
we can see that the DPH self-energy only gives a non-negligble
contribution in the region of parameter space where the coupling to
the leads is small, and at low bias.

This has potential implications for real molecular junctions.
Consider a junction which has its dominant molecular levels far from
the equilibrium Fermi levels of the leads.  At sufficiently high bias,
the DPH self-energy will have a significant contribution to the
inelastic spectra.  This could occur in the case of a junction formed
from an organic molecule if the electronic level of the molecule is
within range of the intended operational bias of the junction.  If
however the dominant molecular electronic level is close to the leads'
Fermi levels, and the coupling to the leads is large (as would be the
case, for example, in a gold nanoconstriction) then DPH will not give
a significant contribution at any applied bias.

\section{Conclusions}
\label{sec:Conclusions}

By using the non-equilibrium Green's functions technique, we have
studied the effect of electron-vibron interaction on the inelastic
transport properties of single-molecule nanojunctions for a model
system.  We have included not only the first-order diagrams (BA) but
also the second-order diagrams (double-exchange DX and dressed phonon
DPH diagrams) for the electron-vibration interaction.  We have
calculated the inelastic electron tunneling spectrum (IETS) across the
full range of parameters available to our model.  The effects of the
second-order DX and DPH diagrams are different and affect different
features of the IETS signal. The effects of these diagrams are
generally less visible in the integrated quantities, such as the
current or the derivated IETS signal, than in the spectral functions
\cite{Dash:2010}.  However their effects are non-negligible, and are
important for the full understanding of the spectroscopic information
conveyed by the IETS signal.

The effect of the dressed-phonon (DPH) diagram is more important in
the bias regions where one of the leads' chemical potentials begins to
impinge upon the electron resonance or one of its vibron replica
(i.e. for resonant inelastic features).  Its effect is reduced both by
increasing the lead-molecule-lead coupling and/or reducing the
electron-vibron interaction.  The renormalization of the vibron
propagator (DPH) has been shown to be strongly dependent on the
self-consistency of the calculations. It would be interesting now to
study the effects of the full series of the electron-hole bubble on
the renormalised vibron propagator (i.e. full $GW$-like diagram).

The double-exchange diagram (DX) affects all the features in the IETS
signal (i.e.  resonant inelastic features and purely inelastic
features at $V \sim\omega_0$).  The corrections are small in the
weak-to-medium \elvib coupling because they are of the order of
$\mathcal{O}(\gamma_0^4)$. However we have shown, numerically and
analytically, that the effect of DX is similar to a dynamical
renormalization of one vertex in the Fock-like diagram.  More
interestingly, the complex form of the non-equilibrium dynamical
renormalized electron-vibron coupling
$\tilde{\gamma}_0(\omega,\omega')$ we have derived analytically can be
adequately replaced in our IETS calculations by a single static
renormalized parameter $\bar{\gamma}_0$. This important result leads
us to believe that the second-order DX calculations, which are extremely
costly in computing time even for our model system, can be
incorporated in calculations for realistic systems by an appropriate
renormalization of the vertex in a low-computing-cost SCBA
calculation.

\appendix
\section{Non-equilibrium vertex corrections to the Fock diagram}
\label{sec:Appendix-vertex}

In Appendix A of Ref.[\onlinecite{Dash:2010}] we have given all the
details for the derivations of the first- and second-order
electron-vibron self-energies. In this section we show how the
second-order double-exchange (DX) diagram can be recast in an
effective first-order Fock-like diagram with a renormalized vertex.

We recall that for the SSSM model, the Fock and DX self-energies
defined on the Keldysh contour $C_K$ are given by
\begin{equation}
  \label{eq:FockSE_onCK}
\Sigma^F_\evib(\tau_1,\tau_2)={\rm i}\gamma_0^2\
D_0(\tau_1,\tau_2)\ G(\tau_1,\tau_2) ,
\end{equation}
and
\begin{equation}
  \label{eq:DXSE_onCK}
  \begin{split}
    \Sigma^{DX}_\evib(\tau_1,\tau_2)  = - \frac{\gamma_0^4}{3} 
    \int_{C_K} &  {\rm d}\tau_3 {\rm d}\tau_4 \
    G(\tau_1,\tau_3) D_0(\tau_1,\tau_4) \times \\ 
    & G(\tau_3,\tau_4) D_0(\tau_3,\tau_2) G(\tau_4,\tau_2)  . 
  \end{split}
\end{equation}
The DX self-energy can be rewritten as an effective Fock-like diagram
after introducing a renormalized \elvib coupling parameter
$\tilde\gamma_0(\tau_3,\tau_4;\tau_2)$ and the vertex function
$\Gamma(\tau_3,\tau_4;\tau_2)$:
\begin{equation}
  \label{eq:DXasFockSE_onCK}
  \begin{split}
\Sigma^{DX}_\evib(\tau_1,\tau_2)  =   {\rm i}\gamma_0
\int_{C_K}  {\rm d}\tau_3 {\rm d}\tau_4  & \ \tilde\gamma_0(\tau_3,\tau_4;\tau_2) \times \\
 & D_0(\tau_1,\tau_4) G(\tau_1,\tau_3)  ,
  \end{split}
\end{equation}
with 
\begin{equation}
  \label{eq:renormgamma0_onCK}
\tilde\gamma_0(\tau_3,\tau_4;\tau_2) = \frac{{\rm i} \gamma_0^3}{3} \Gamma(\tau_3,\tau_4;\tau_2),
\end{equation}
and where the vertex function is given by 
\begin{equation}
  \label{eq:vertexfnc_onCK}
\Gamma(\tau_3,\tau_4;\tau_2) = G(\tau_3,\tau_4) D_0(\tau_3,\tau_2) G(\tau_4,\tau_2)  .
\end{equation}

The above expression for the vertex function is compatible with the
second-order expansion of the \elvib interaction.  A generalization of
the vertex function (see Fock-like diagram in Figure
\ref{fig:DX-Feynman-mod}) to all orders of the interaction is
possible, though beyond the scope of the present paper.  Note that at
the lowest order, the renormalized \elvib coupling parameter would be
given by $\tilde\gamma_0(\tau_3,\tau_4;\tau_2)=\gamma_0
\Gamma^{(0)}(\tau_3,\tau_4;\tau_2)$ with
$\Gamma^{(0)}(\tau_3,\tau_4;\tau_2)=
\delta(\tau_3-\tau_2)\delta(\tau_4-\tau_2)$.  Hence
Eq.(\ref{eq:DXSE_onCK}) would simply be transformed
Eq.(\ref{eq:FockSE_onCK}) as expected.

Using the rules of analytical continuation on the real-time branches
given in Appendix A of paper Ref.[\onlinecite{Dash:2010}], we find
the different components of the self-energies. Then after taking the
Fourier transform of the different quantities in the steady-state
limit, i.e. $X(t,t')=X(t-t')$, we find the following expression for
the different components of the energy-dependent self-energies:

\begin{equation}
  \label{eq:FockSE_w}
\Sigma_{\evib}^{F,\zeta_1\zeta_2}(\omega)  =  i \gamma_0^2 
\int \frac{dv}{2\pi} \
    G^{\zeta_1\zeta_2}(v) \ 
    D_0^{\zeta_1\zeta_2}(\omega-v)  ,
\end{equation}
and
\begin{equation}
\label{eq:DXSE_w}
\begin{split}
\Sigma^{DX,\zeta_1\zeta_2}_\evib(\omega)  = {\rm i} \gamma_0
\int \frac{{\rm d}v}{2\pi} 
\sum_{\zeta_3,\zeta_4} \zeta_3 \zeta_4 \ 
& G^{\zeta_1 \zeta_3}(v) 
D_0^{\zeta_1 \zeta_4}(\omega-v) \times \\
& \tilde\gamma_0^{\zeta_3 \zeta_4 \zeta_2}(\omega,v),
\end{split}
\end{equation}
where the non-equilibrium dynamical renormalised \elvib coupling is
given by
\begin{equation}
\label{eq:renormgamma0_wv}
\begin{split}
\tilde\gamma_0^{\zeta_3 \zeta_4 \zeta_2}(\omega,v) =
\frac{{\rm i} \gamma_0^3}{3}
\int \frac{{\rm d}u}{2\pi} \
& G^{\zeta_3 \zeta_4}(v-u) \times \\
& D_0^{\zeta_3 \zeta_2}(u) G^{\zeta_4 \zeta_2}(\omega-u)  .
\end{split}
\end{equation}

(The index $\zeta_i=\pm$ labels the branch of the Keldysh time-loop
contour $C_K$ and are related to the usual convention: time-ordered
($t=++$), anti time-ordered ($\tilde t=--$), greater ($>=-+$) and
lesser ($<=+-$) components.)

Because all the quantities are originally defined on $C_K$ and because
the vertex function is a 3-point (3 times) function, the
non-equilibrium dynamical renormalized \elvib coupling is a complex
function of three indices $\zeta_i$ and of two energy variables.  Such
a dynamical renormalization (including non-equilibrium conditions) is
much more complicated than a simple static renormalization of the \elvib
constant coupling $\gamma_0 \rightarrow \tilde\gamma_0$.

In our analysis of the IETS signal in the off-resonant
transport regime, we try to keep the interpretation of the results as
simple as possible, and we show that the renormalization of the IETS
signal due to the DX diagram can be fairly well approximated by a
simple static renormalization of the coupling constant
$\gamma_0$ for applied bias around the vibration frequency $V\sim
\omega_0 \pm 20 \%$.

\begin{acknowledgments}
This work was funded in part by the European Community's Seventh
Framework Programme (FP7/2007-2013) under grant agreement no 211956
(ETSF e-I3 grant).
\end{acknowledgments} 


\begin{thebibliography}{63}
\expandafter\ifx\csname natexlab\endcsname\relax\def\natexlab#1{#1}\fi
\expandafter\ifx\csname bibnamefont\endcsname\relax
  \def\bibnamefont#1{#1}\fi
\expandafter\ifx\csname bibfnamefont\endcsname\relax
  \def\bibfnamefont#1{#1}\fi
\expandafter\ifx\csname citenamefont\endcsname\relax
  \def\citenamefont#1{#1}\fi
\expandafter\ifx\csname url\endcsname\relax
  \def\url#1{\texttt{#1}}\fi
\expandafter\ifx\csname urlprefix\endcsname\relax\def\urlprefix{URL }\fi
\providecommand{\bibinfo}[2]{#2}
\providecommand{\eprint}[2][]{\url{#2}}

\bibitem[{\citenamefont{Hipps and Mazur}(1993)}]{Hipps:1993}
\bibinfo{author}{\bibfnamefont{K.~W.} \bibnamefont{Hipps}} \bibnamefont{and}
  \bibinfo{author}{\bibfnamefont{U.}~\bibnamefont{Mazur}},
  \bibinfo{journal}{Journal of Physical Chemistry}
  \textbf{\bibinfo{volume}{97}}, \bibinfo{pages}{7803} (\bibinfo{year}{1993}).

\bibitem[{\citenamefont{Liu et~al.}(2004)\citenamefont{Liu, Pradhan, and
  Ho}}]{Liu:2004}
\bibinfo{author}{\bibfnamefont{N.}~\bibnamefont{Liu}},
  \bibinfo{author}{\bibfnamefont{N.~A.} \bibnamefont{Pradhan}},
  \bibnamefont{and} \bibinfo{author}{\bibfnamefont{W.}~\bibnamefont{Ho}},
  \bibinfo{journal}{Journal of Chemical Physics}
  \textbf{\bibinfo{volume}{120}}, \bibinfo{pages}{11371}
  (\bibinfo{year}{2004}).

\bibitem[{\citenamefont{Kushmerick et~al.}(2004)\citenamefont{Kushmerick,
  Lazorcik, Patterson, and Shashidhar}}]{Kushmerick:2004}
\bibinfo{author}{\bibfnamefont{J.~G.} \bibnamefont{Kushmerick}},
  \bibinfo{author}{\bibfnamefont{J.}~\bibnamefont{Lazorcik}},
  \bibinfo{author}{\bibfnamefont{C.~H.} \bibnamefont{Patterson}},
  \bibnamefont{and}
  \bibinfo{author}{\bibfnamefont{R.}~\bibnamefont{Shashidhar}},
  \bibinfo{journal}{Nano Letters} \textbf{\bibinfo{volume}{4}},
  \bibinfo{pages}{639} (\bibinfo{year}{2004}).

\bibitem[{\citenamefont{Yu et~al.}(2004)\citenamefont{Yu, Keane, Ciszek, Cheng,
  Stewart, Tour, and Natelson}}]{Yu:2004}
\bibinfo{author}{\bibfnamefont{L.~H.} \bibnamefont{Yu}},
  \bibinfo{author}{\bibfnamefont{Z.~K.} \bibnamefont{Keane}},
  \bibinfo{author}{\bibfnamefont{J.~W.} \bibnamefont{Ciszek}},
  \bibinfo{author}{\bibfnamefont{L.}~\bibnamefont{Cheng}},
  \bibinfo{author}{\bibfnamefont{M.~P.} \bibnamefont{Stewart}},
  \bibinfo{author}{\bibfnamefont{J.~M.} \bibnamefont{Tour}}, \bibnamefont{and}
  \bibinfo{author}{\bibfnamefont{D.}~\bibnamefont{Natelson}},
  \bibinfo{journal}{Physical Review Letters} \textbf{\bibinfo{volume}{93}},
  \bibinfo{pages}{266802} (\bibinfo{year}{2004}).

\bibitem[{\citenamefont{Yu et~al.}(2006)\citenamefont{Yu, Zangmeister, and
  Kushmerick}}]{Yu:2006}
\bibinfo{author}{\bibfnamefont{L.~H.} \bibnamefont{Yu}},
  \bibinfo{author}{\bibfnamefont{C.~D.} \bibnamefont{Zangmeister}},
  \bibnamefont{and} \bibinfo{author}{\bibfnamefont{J.~G.}
  \bibnamefont{Kushmerick}}, \bibinfo{journal}{Nano Letters}
  \textbf{\bibinfo{volume}{6}}, \bibinfo{pages}{2515} (\bibinfo{year}{2006}).

\bibitem[{\citenamefont{Chae et~al.}(2006)\citenamefont{Chae, Berry, Jung,
  Murillo, and Yao}}]{Chae:2006}
\bibinfo{author}{\bibfnamefont{D.-H.} \bibnamefont{Chae}},
  \bibinfo{author}{\bibfnamefont{J.~F.} \bibnamefont{Berry}},
  \bibinfo{author}{\bibfnamefont{S.}~\bibnamefont{Jung}},
  \bibinfo{author}{\bibfnamefont{F.~A. C. C.~A.} \bibnamefont{Murillo}},
  \bibnamefont{and} \bibinfo{author}{\bibfnamefont{Z.}~\bibnamefont{Yao}},
  \bibinfo{journal}{Nano Letters} \textbf{\bibinfo{volume}{6}},
  \bibinfo{pages}{165} (\bibinfo{year}{2006}).

\bibitem[{\citenamefont{Beebe et~al.}(2007)\citenamefont{Beebe, Moore, Lee, and
  Kushmerick}}]{Beebe_JM:2007}
\bibinfo{author}{\bibfnamefont{J.~M.} \bibnamefont{Beebe}},
  \bibinfo{author}{\bibfnamefont{H.~J.} \bibnamefont{Moore}},
  \bibinfo{author}{\bibfnamefont{T.~R.} \bibnamefont{Lee}}, \bibnamefont{and}
  \bibinfo{author}{\bibfnamefont{J.~G.} \bibnamefont{Kushmerick}},
  \bibinfo{journal}{Nano Letters} \textbf{\bibinfo{volume}{7}},
  \bibinfo{pages}{1364} (\bibinfo{year}{2007}).

\bibitem[{\citenamefont{Okabayashi et~al.}(2008)\citenamefont{Okabayashi,
  Konda, and Komeda}}]{Okabayashi:2008}
\bibinfo{author}{\bibfnamefont{N.}~\bibnamefont{Okabayashi}},
  \bibinfo{author}{\bibfnamefont{Y.}~\bibnamefont{Konda}}, \bibnamefont{and}
  \bibinfo{author}{\bibfnamefont{T.}~\bibnamefont{Komeda}},
  \bibinfo{journal}{Physical Review Letters} \textbf{\bibinfo{volume}{100}},
  \bibinfo{pages}{217801} (\bibinfo{year}{2008}).

\bibitem[{\citenamefont{Gawronski et~al.}(2009)\citenamefont{Gawronski,
  Fransson, and Morgenstern}}]{Gawronski:2009}
\bibinfo{author}{\bibfnamefont{H.}~\bibnamefont{Gawronski}},
  \bibinfo{author}{\bibfnamefont{J.}~\bibnamefont{Fransson}}, \bibnamefont{and}
  \bibinfo{author}{\bibfnamefont{K.}~\bibnamefont{Morgenstern}},
  \emph{\bibinfo{title}{Imaging of inelastic waves in iets maps}}
  (\bibinfo{year}{2009}), \eprint[arXiv]{arXiv:0911.4053v1}.

\bibitem[{\citenamefont{Kim et~al.}()\citenamefont{Kim, Song, Strigl, Pernau,
  and and}}]{Kim:2010}
\bibinfo{author}{\bibfnamefont{Y.}~\bibnamefont{Kim}},
  \bibinfo{author}{\bibfnamefont{H.}~\bibnamefont{Song}},
  \bibinfo{author}{\bibfnamefont{F.}~\bibnamefont{Strigl}},
  \bibinfo{author}{\bibfnamefont{H.-F.} \bibnamefont{Pernau}},
  \bibnamefont{and} \bibinfo{author}{\bibfnamefont{T.~L.} \bibnamefont{and}},
  \emph{\bibinfo{title}{Mechanical control of vibrational states in
  single-molecule junctions}}, \eprint[arXiv]{arXiv:1011.3226v1}.

\bibitem[{\citenamefont{Hyldgaard et~al.}(1994)\citenamefont{Hyldgaard,
  Hershfield, Davies, and Wilkins}}]{Hyldgaard:1994}
\bibinfo{author}{\bibfnamefont{P.}~\bibnamefont{Hyldgaard}},
  \bibinfo{author}{\bibfnamefont{S.}~\bibnamefont{Hershfield}},
  \bibinfo{author}{\bibfnamefont{J.~H.} \bibnamefont{Davies}},
  \bibnamefont{and} \bibinfo{author}{\bibfnamefont{J.~W.}
  \bibnamefont{Wilkins}}, \bibinfo{journal}{Annals of Physics}
  \textbf{\bibinfo{volume}{236}}, \bibinfo{pages}{1} (\bibinfo{year}{1994}).

\bibitem[{\citenamefont{Ness and Fisher}(1999)}]{Ness:1999}
\bibinfo{author}{\bibfnamefont{H.}~\bibnamefont{Ness}} \bibnamefont{and}
  \bibinfo{author}{\bibfnamefont{A.~J.} \bibnamefont{Fisher}},
  \bibinfo{journal}{Physical Review Letters} \textbf{\bibinfo{volume}{83}},
  \bibinfo{pages}{452} (\bibinfo{year}{1999}).

\bibitem[{\citenamefont{Ness et~al.}(2001)\citenamefont{Ness, Shevlin, and
  Fisher}}]{Ness:2001}
\bibinfo{author}{\bibfnamefont{H.}~\bibnamefont{Ness}},
  \bibinfo{author}{\bibfnamefont{S.~A.} \bibnamefont{Shevlin}},
  \bibnamefont{and} \bibinfo{author}{\bibfnamefont{A.~J.}
  \bibnamefont{Fisher}}, \bibinfo{journal}{Physical Review B}
  \textbf{\bibinfo{volume}{63}}, \bibinfo{pages}{125422}
  (\bibinfo{year}{2001}).

\bibitem[{\citenamefont{Ness and Fisher}(2002)}]{Ness:2002a}
\bibinfo{author}{\bibfnamefont{H.}~\bibnamefont{Ness}} \bibnamefont{and}
  \bibinfo{author}{\bibfnamefont{A.~J.} \bibnamefont{Fisher}},
  \bibinfo{journal}{Europhysics Letters} \textbf{\bibinfo{volume}{57}},
  \bibinfo{pages}{885} (\bibinfo{year}{2002}).

\bibitem[{\citenamefont{Flensberg}(2003)}]{Flensberg:2003}
\bibinfo{author}{\bibfnamefont{K.}~\bibnamefont{Flensberg}},
  \bibinfo{journal}{Physical Review B} \textbf{\bibinfo{volume}{68}},
  \bibinfo{pages}{205323} (\bibinfo{year}{2003}).

\bibitem[{\citenamefont{Mii et~al.}(2003)\citenamefont{Mii, Tikhodeev, and
  Ueba}}]{Mii:2003}
\bibinfo{author}{\bibfnamefont{T.}~\bibnamefont{Mii}},
  \bibinfo{author}{\bibfnamefont{S.}~\bibnamefont{Tikhodeev}},
  \bibnamefont{and} \bibinfo{author}{\bibfnamefont{H.}~\bibnamefont{Ueba}},
  \bibinfo{journal}{Physical Review B} \textbf{\bibinfo{volume}{68}},
  \bibinfo{pages}{205406} (\bibinfo{year}{2003}).

\bibitem[{\citenamefont{Montgomery et~al.}(2003)\citenamefont{Montgomery,
  Hoekstra, Sutton, and Todorov}}]{Montgomery:2003b}
\bibinfo{author}{\bibfnamefont{M.~J.} \bibnamefont{Montgomery}},
  \bibinfo{author}{\bibfnamefont{J.}~\bibnamefont{Hoekstra}},
  \bibinfo{author}{\bibfnamefont{A.~P.} \bibnamefont{Sutton}},
  \bibnamefont{and} \bibinfo{author}{\bibfnamefont{T.~N.}
  \bibnamefont{Todorov}}, \bibinfo{journal}{Journal of Physics: Condensed
  Matter} \textbf{\bibinfo{volume}{15}}, \bibinfo{pages}{731}
  (\bibinfo{year}{2003}).

\bibitem[{\citenamefont{Troisi et~al.}(2003)\citenamefont{Troisi, Ratner, and
  Nitzan}}]{Troisi:2003}
\bibinfo{author}{\bibfnamefont{A.}~\bibnamefont{Troisi}},
  \bibinfo{author}{\bibfnamefont{M.~A.} \bibnamefont{Ratner}},
  \bibnamefont{and} \bibinfo{author}{\bibfnamefont{A.}~\bibnamefont{Nitzan}},
  \bibinfo{journal}{Journal of Chemical Physics}
  \textbf{\bibinfo{volume}{118}}, \bibinfo{pages}{6072} (\bibinfo{year}{2003}).

\bibitem[{\citenamefont{Chen et~al.}(2005{\natexlab{a}})\citenamefont{Chen,
  Zwolak, and di~Ventra}}]{Chen:2004}
\bibinfo{author}{\bibfnamefont{Y.~C.} \bibnamefont{Chen}},
  \bibinfo{author}{\bibfnamefont{M.}~\bibnamefont{Zwolak}}, \bibnamefont{and}
  \bibinfo{author}{\bibfnamefont{M.}~\bibnamefont{di~Ventra}},
  \bibinfo{journal}{Nano Letters} \textbf{\bibinfo{volume}{4}},
  \bibinfo{pages}{1709} (\bibinfo{year}{2005}{\natexlab{a}}).

\bibitem[{\citenamefont{Lorente and Persson}(2000)}]{Lorente:2000}
\bibinfo{author}{\bibfnamefont{N.}~\bibnamefont{Lorente}} \bibnamefont{and}
  \bibinfo{author}{\bibfnamefont{M.}~\bibnamefont{Persson}},
  \bibinfo{journal}{Physical Review Letters} \textbf{\bibinfo{volume}{85}},
  \bibinfo{pages}{2997} (\bibinfo{year}{2000}).

\bibitem[{\citenamefont{Frederiksen et~al.}(2004)\citenamefont{Frederiksen,
  Brandbyge, Lorente, and Jauho}}]{Frederiksen:2004}
\bibinfo{author}{\bibfnamefont{T.}~\bibnamefont{Frederiksen}},
  \bibinfo{author}{\bibfnamefont{M.}~\bibnamefont{Brandbyge}},
  \bibinfo{author}{\bibfnamefont{N.}~\bibnamefont{Lorente}}, \bibnamefont{and}
  \bibinfo{author}{\bibfnamefont{A.~P.} \bibnamefont{Jauho}},
  \bibinfo{journal}{Physical Review Letters} \textbf{\bibinfo{volume}{93}},
  \bibinfo{pages}{256601} (\bibinfo{year}{2004}).

\bibitem[{\citenamefont{Galperin
  et~al.}(2004{\natexlab{a}})\citenamefont{Galperin, Ratner, and
  Nitzan}}]{Galperin:2004}
\bibinfo{author}{\bibfnamefont{M.}~\bibnamefont{Galperin}},
  \bibinfo{author}{\bibfnamefont{M.~A.} \bibnamefont{Ratner}},
  \bibnamefont{and} \bibinfo{author}{\bibfnamefont{A.}~\bibnamefont{Nitzan}},
  \bibinfo{journal}{Nano Letters} \textbf{\bibinfo{volume}{4}},
  \bibinfo{pages}{1605} (\bibinfo{year}{2004}{\natexlab{a}}).

\bibitem[{\citenamefont{Galperin
  et~al.}(2004{\natexlab{b}})\citenamefont{Galperin, Ratner, and
  Nitzan}}]{Galperin:2004b}
\bibinfo{author}{\bibfnamefont{M.}~\bibnamefont{Galperin}},
  \bibinfo{author}{\bibfnamefont{M.~A.} \bibnamefont{Ratner}},
  \bibnamefont{and} \bibinfo{author}{\bibfnamefont{A.}~\bibnamefont{Nitzan}},
  \bibinfo{journal}{Journal of Chemical Physics}
  \textbf{\bibinfo{volume}{121}}, \bibinfo{pages}{11965}
  (\bibinfo{year}{2004}{\natexlab{b}}).

\bibitem[{\citenamefont{Mitra et~al.}(2004)\citenamefont{Mitra, Aleiner, and
  Millis}}]{Mitra:2004}
\bibinfo{author}{\bibfnamefont{A.}~\bibnamefont{Mitra}},
  \bibinfo{author}{\bibfnamefont{I.}~\bibnamefont{Aleiner}}, \bibnamefont{and}
  \bibinfo{author}{\bibfnamefont{A.~J.} \bibnamefont{Millis}},
  \bibinfo{journal}{Physical Review B} \textbf{\bibinfo{volume}{69}},
  \bibinfo{pages}{245302} (\bibinfo{year}{2004}).

\bibitem[{\citenamefont{Pecchia et~al.}(2004)\citenamefont{Pecchia, di~Carlo,
  Gagliardi, Sanna, Frauenhein, and Gutierrez}}]{Pecchia:2004}
\bibinfo{author}{\bibfnamefont{A.}~\bibnamefont{Pecchia}},
  \bibinfo{author}{\bibfnamefont{A.}~\bibnamefont{di~Carlo}},
  \bibinfo{author}{\bibfnamefont{A.}~\bibnamefont{Gagliardi}},
  \bibinfo{author}{\bibfnamefont{S.}~\bibnamefont{Sanna}},
  \bibinfo{author}{\bibfnamefont{T.}~\bibnamefont{Frauenhein}},
  \bibnamefont{and}
  \bibinfo{author}{\bibfnamefont{R.}~\bibnamefont{Gutierrez}},
  \bibinfo{journal}{Nano Letters} \textbf{\bibinfo{volume}{4}},
  \bibinfo{pages}{2109} (\bibinfo{year}{2004}).

\bibitem[{\citenamefont{Pecchia and di~Carlo}(2004)}]{Pecchia:2004b}
\bibinfo{author}{\bibfnamefont{A.}~\bibnamefont{Pecchia}} \bibnamefont{and}
  \bibinfo{author}{\bibfnamefont{A.}~\bibnamefont{di~Carlo}},
  \bibinfo{journal}{Reports on Progress in Physics}
  \textbf{\bibinfo{volume}{67}}, \bibinfo{pages}{1497} (\bibinfo{year}{2004}).

\bibitem[{\citenamefont{Chen et~al.}(2005{\natexlab{b}})\citenamefont{Chen,
  L\"u, and Zhu}}]{Chen_Z:2005}
\bibinfo{author}{\bibfnamefont{Z.}~\bibnamefont{Chen}},
  \bibinfo{author}{\bibfnamefont{R.}~\bibnamefont{L\"u}}, \bibnamefont{and}
  \bibinfo{author}{\bibfnamefont{B.}~\bibnamefont{Zhu}},
  \bibinfo{journal}{Physical Review B} \textbf{\bibinfo{volume}{71}},
  \bibinfo{pages}{165324} (\bibinfo{year}{2005}{\natexlab{b}}).

\bibitem[{\citenamefont{Paulsson et~al.}(2005)\citenamefont{Paulsson,
  Frederiksen, and Brandbyge}}]{Paulsson:2005}
\bibinfo{author}{\bibfnamefont{M.}~\bibnamefont{Paulsson}},
  \bibinfo{author}{\bibfnamefont{T.}~\bibnamefont{Frederiksen}},
  \bibnamefont{and}
  \bibinfo{author}{\bibfnamefont{M.}~\bibnamefont{Brandbyge}},
  \bibinfo{journal}{Physical Review B} \textbf{\bibinfo{volume}{72}},
  \bibinfo{pages}{201101} (\bibinfo{year}{2005}).

\bibitem[{\citenamefont{Ryndyk and Keller}(2005)}]{Ryndyk:2005}
\bibinfo{author}{\bibfnamefont{D.~A.} \bibnamefont{Ryndyk}} \bibnamefont{and}
  \bibinfo{author}{\bibfnamefont{J.}~\bibnamefont{Keller}},
  \bibinfo{journal}{Physical Review B} \textbf{\bibinfo{volume}{71}},
  \bibinfo{pages}{073305} (\bibinfo{year}{2005}).

\bibitem[{\citenamefont{Sergueev et~al.}(2005)\citenamefont{Sergueev, Roubtsov,
  and Guo}}]{Sergueev:2005}
\bibinfo{author}{\bibfnamefont{N.}~\bibnamefont{Sergueev}},
  \bibinfo{author}{\bibfnamefont{D.}~\bibnamefont{Roubtsov}}, \bibnamefont{and}
  \bibinfo{author}{\bibfnamefont{H.}~\bibnamefont{Guo}},
  \bibinfo{journal}{Physical Review Letters} \textbf{\bibinfo{volume}{95}},
  \bibinfo{pages}{146803} (\bibinfo{year}{2005}).

\bibitem[{\citenamefont{Viljas et~al.}(2005)\citenamefont{Viljas, Cuevas,
  Pauly, and H\"afner}}]{Viljas:2005}
\bibinfo{author}{\bibfnamefont{J.~K.} \bibnamefont{Viljas}},
  \bibinfo{author}{\bibfnamefont{J.~C.} \bibnamefont{Cuevas}},
  \bibinfo{author}{\bibfnamefont{F.}~\bibnamefont{Pauly}}, \bibnamefont{and}
  \bibinfo{author}{\bibfnamefont{M.}~\bibnamefont{H\"afner}},
  \bibinfo{journal}{Physical Review B} \textbf{\bibinfo{volume}{72}},
  \bibinfo{pages}{245415} (\bibinfo{year}{2005}).

\bibitem[{\citenamefont{Yamamoto et~al.}(2005)\citenamefont{Yamamoto, Watanabe,
  and Watanabe}}]{Yamamoto:2005}
\bibinfo{author}{\bibfnamefont{T.}~\bibnamefont{Yamamoto}},
  \bibinfo{author}{\bibfnamefont{K.}~\bibnamefont{Watanabe}}, \bibnamefont{and}
  \bibinfo{author}{\bibfnamefont{S.}~\bibnamefont{Watanabe}},
  \bibinfo{journal}{Physical Review Letters} \textbf{\bibinfo{volume}{95}},
  \bibinfo{pages}{065501} (\bibinfo{year}{2005}).

\bibitem[{\citenamefont{Cresti et~al.}(2006)\citenamefont{Cresti, Grosso, and
  Parravicini}}]{Cresti:2006}
\bibinfo{author}{\bibfnamefont{A.}~\bibnamefont{Cresti}},
  \bibinfo{author}{\bibfnamefont{G.}~\bibnamefont{Grosso}}, \bibnamefont{and}
  \bibinfo{author}{\bibfnamefont{G.~P.} \bibnamefont{Parravicini}},
  \bibinfo{journal}{Journal of Physics: Condensed Matter}
  \textbf{\bibinfo{volume}{18}}, \bibinfo{pages}{10059} (\bibinfo{year}{2006}).

\bibitem[{\citenamefont{Kula et~al.}(2006)\citenamefont{Kula, Jiang, and
  Luo}}]{Kula:2006}
\bibinfo{author}{\bibfnamefont{M.}~\bibnamefont{Kula}},
  \bibinfo{author}{\bibfnamefont{J.}~\bibnamefont{Jiang}}, \bibnamefont{and}
  \bibinfo{author}{\bibfnamefont{Y.}~\bibnamefont{Luo}}, \bibinfo{journal}{Nano
  Letters} \textbf{\bibinfo{volume}{6}}, \bibinfo{pages}{1693}
  (\bibinfo{year}{2006}).

\bibitem[{\citenamefont{Paulsson et~al.}(2006)\citenamefont{Paulsson,
  Frederiksen, and Brandbyge}}]{Paulsson:2006}
\bibinfo{author}{\bibfnamefont{M.}~\bibnamefont{Paulsson}},
  \bibinfo{author}{\bibfnamefont{T.}~\bibnamefont{Frederiksen}},
  \bibnamefont{and}
  \bibinfo{author}{\bibfnamefont{M.}~\bibnamefont{Brandbyge}},
  \bibinfo{journal}{Nano Letters} \textbf{\bibinfo{volume}{6}},
  \bibinfo{pages}{258} (\bibinfo{year}{2006}).

\bibitem[{\citenamefont{Ryndyk et~al.}(2006)\citenamefont{Ryndyk, Hartung, and
  Cuniberti}}]{Ryndyk:2006}
\bibinfo{author}{\bibfnamefont{D.~A.} \bibnamefont{Ryndyk}},
  \bibinfo{author}{\bibfnamefont{M.}~\bibnamefont{Hartung}}, \bibnamefont{and}
  \bibinfo{author}{\bibfnamefont{G.}~\bibnamefont{Cuniberti}},
  \bibinfo{journal}{Physical Review B} \textbf{\bibinfo{volume}{73}},
  \bibinfo{pages}{045420} (\bibinfo{year}{2006}).

\bibitem[{\citenamefont{Troisi and Ratner}(2006)}]{Troisi:2006b}
\bibinfo{author}{\bibfnamefont{A.}~\bibnamefont{Troisi}} \bibnamefont{and}
  \bibinfo{author}{\bibfnamefont{M.~A.} \bibnamefont{Ratner}},
  \bibinfo{journal}{Nano Letters} \textbf{\bibinfo{volume}{6}},
  \bibinfo{pages}{1784} (\bibinfo{year}{2006}).

\bibitem[{\citenamefont{de~la Vega et~al.}(2006)\citenamefont{de~la Vega,
  Mart\'{\i}n-Rodero, Agra\"{\i}t, and Levy-Yeyati}}]{Vega:2006}
\bibinfo{author}{\bibfnamefont{L.}~\bibnamefont{de~la Vega}},
  \bibinfo{author}{\bibfnamefont{A.}~\bibnamefont{Mart\'{\i}n-Rodero}},
  \bibinfo{author}{\bibfnamefont{N.}~\bibnamefont{Agra\"{\i}t}},
  \bibnamefont{and}
  \bibinfo{author}{\bibfnamefont{A.}~\bibnamefont{Levy-Yeyati}},
  \bibinfo{journal}{Physical Review B} \textbf{\bibinfo{volume}{73}},
  \bibinfo{pages}{075428} (\bibinfo{year}{2006}).

\bibitem[{\citenamefont{Toroker and Peskin}(2007)}]{Caspary:2007}
\bibinfo{author}{\bibfnamefont{M.~C.} \bibnamefont{Toroker}} \bibnamefont{and}
  \bibinfo{author}{\bibfnamefont{U.}~\bibnamefont{Peskin}},
  \bibinfo{journal}{Journal of Chemical Physics}
  \textbf{\bibinfo{volume}{127}}, \bibinfo{pages}{154706}
  (\bibinfo{year}{2007}).

\bibitem[{\citenamefont{Frederiksen et~al.}(2007)\citenamefont{Frederiksen,
  Paulsson, Brandbyge, and Jauho}}]{Frederiksen:2007}
\bibinfo{author}{\bibfnamefont{T.}~\bibnamefont{Frederiksen}},
  \bibinfo{author}{\bibfnamefont{M.}~\bibnamefont{Paulsson}},
  \bibinfo{author}{\bibfnamefont{M.}~\bibnamefont{Brandbyge}},
  \bibnamefont{and} \bibinfo{author}{\bibfnamefont{A.-P.} \bibnamefont{Jauho}},
  \bibinfo{journal}{Physical Review B} \textbf{\bibinfo{volume}{75}},
  \bibinfo{pages}{205413} (\bibinfo{year}{2007}).

\bibitem[{\citenamefont{Galperin et~al.}(2007)\citenamefont{Galperin, Nitzan,
  and Ratner}}]{Galperin:2007}
\bibinfo{author}{\bibfnamefont{M.}~\bibnamefont{Galperin}},
  \bibinfo{author}{\bibfnamefont{A.}~\bibnamefont{Nitzan}}, \bibnamefont{and}
  \bibinfo{author}{\bibfnamefont{M.~A.} \bibnamefont{Ratner}},
  \bibinfo{journal}{Physical Review B} \textbf{\bibinfo{volume}{74}},
  \bibinfo{pages}{075326} (\bibinfo{year}{2007}).

\bibitem[{\citenamefont{Ryndyk and Cuniberti}(2007)}]{Ryndyk:2007}
\bibinfo{author}{\bibfnamefont{D.~A.} \bibnamefont{Ryndyk}} \bibnamefont{and}
  \bibinfo{author}{\bibfnamefont{G.}~\bibnamefont{Cuniberti}},
  \bibinfo{journal}{Physical Review B} \textbf{\bibinfo{volume}{76}},
  \bibinfo{pages}{155430} (\bibinfo{year}{2007}).

\bibitem[{\citenamefont{Schmidt et~al.}(2007)\citenamefont{Schmidt, Hettler,
  and Sch\"on}}]{Schmidt:2007}
\bibinfo{author}{\bibfnamefont{B.~B.} \bibnamefont{Schmidt}},
  \bibinfo{author}{\bibfnamefont{M.~H.} \bibnamefont{Hettler}},
  \bibnamefont{and} \bibinfo{author}{\bibfnamefont{G.}~\bibnamefont{Sch\"on}},
  \bibinfo{journal}{Physical Review B} \textbf{\bibinfo{volume}{75}},
  \bibinfo{pages}{115125} (\bibinfo{year}{2007}).

\bibitem[{\citenamefont{Troisi et~al.}(2007)\citenamefont{Troisi, Beebe,
  Picraux, van Zee, Stewart, Ratner, and Kushmerick}}]{Troisi:2007}
\bibinfo{author}{\bibfnamefont{A.}~\bibnamefont{Troisi}},
  \bibinfo{author}{\bibfnamefont{J.~M.} \bibnamefont{Beebe}},
  \bibinfo{author}{\bibfnamefont{L.~B.} \bibnamefont{Picraux}},
  \bibinfo{author}{\bibfnamefont{R.~D.} \bibnamefont{van Zee}},
  \bibinfo{author}{\bibfnamefont{D.~R.} \bibnamefont{Stewart}},
  \bibinfo{author}{\bibfnamefont{M.~A.} \bibnamefont{Ratner}},
  \bibnamefont{and} \bibinfo{author}{\bibfnamefont{J.~G.}
  \bibnamefont{Kushmerick}}, \bibinfo{journal}{Proceedings of the National
  Academy of Sciences of the USA} \textbf{\bibinfo{volume}{104}},
  \bibinfo{pages}{14255} (\bibinfo{year}{2007}).

\bibitem[{\citenamefont{Asai}(2008)}]{Asai:2008}
\bibinfo{author}{\bibfnamefont{Y.}~\bibnamefont{Asai}},
  \bibinfo{journal}{Physical Review B} \textbf{\bibinfo{volume}{78}},
  \bibinfo{pages}{045434} (\bibinfo{year}{2008}).

\bibitem[{\citenamefont{Benesch et~al.}(2008)\citenamefont{Benesch,
  \v{C}\'{\i}\v{z}ek, Klime\v{s}, Kondov, Thoss, and Domcke}}]{Benesch:2008}
\bibinfo{author}{\bibfnamefont{C.}~\bibnamefont{Benesch}},
  \bibinfo{author}{\bibfnamefont{M.}~\bibnamefont{\v{C}\'{\i}\v{z}ek}},
  \bibinfo{author}{\bibfnamefont{J.}~\bibnamefont{Klime\v{s}}},
  \bibinfo{author}{\bibfnamefont{I.}~\bibnamefont{Kondov}},
  \bibinfo{author}{\bibfnamefont{M.}~\bibnamefont{Thoss}}, \bibnamefont{and}
  \bibinfo{author}{\bibfnamefont{W.}~\bibnamefont{Domcke}},
  \bibinfo{journal}{Journal of Physical Chemistry C}
  \textbf{\bibinfo{volume}{112}}, \bibinfo{pages}{9880} (\bibinfo{year}{2008}).

\bibitem[{\citenamefont{Paulsson et~al.}(2008)\citenamefont{Paulsson,
  Frederiksen, Ueba, Lorente, and Brandbyge}}]{Paulsson:2008}
\bibinfo{author}{\bibfnamefont{M.}~\bibnamefont{Paulsson}},
  \bibinfo{author}{\bibfnamefont{T.}~\bibnamefont{Frederiksen}},
  \bibinfo{author}{\bibfnamefont{H.}~\bibnamefont{Ueba}},
  \bibinfo{author}{\bibfnamefont{N.}~\bibnamefont{Lorente}}, \bibnamefont{and}
  \bibinfo{author}{\bibfnamefont{M.}~\bibnamefont{Brandbyge}},
  \bibinfo{journal}{Physical Review Letters} \textbf{\bibinfo{volume}{100}},
  \bibinfo{pages}{226604} (\bibinfo{year}{2008}).

\bibitem[{\citenamefont{Egger and Gogolin}(2008)}]{Egger:2008}
\bibinfo{author}{\bibfnamefont{R.}~\bibnamefont{Egger}} \bibnamefont{and}
  \bibinfo{author}{\bibfnamefont{A.~O.} \bibnamefont{Gogolin}},
  \bibinfo{journal}{Physical Review B} \textbf{\bibinfo{volume}{77}},
  \bibinfo{pages}{113405} (\bibinfo{year}{2008}).

\bibitem[{\citenamefont{Monturet and Lorente}(2008)}]{Monturet:2008}
\bibinfo{author}{\bibfnamefont{S.}~\bibnamefont{Monturet}} \bibnamefont{and}
  \bibinfo{author}{\bibfnamefont{N.}~\bibnamefont{Lorente}},
  \bibinfo{journal}{Physical Review B} \textbf{\bibinfo{volume}{78}},
  \bibinfo{pages}{035445} (\bibinfo{year}{2008}).

\bibitem[{\citenamefont{McEniry et~al.}(2008)\citenamefont{McEniry,
  Frederiksen, Todorov, Dundas, and Horsfield}}]{McEniry:2008}
\bibinfo{author}{\bibfnamefont{E.~J.} \bibnamefont{McEniry}},
  \bibinfo{author}{\bibfnamefont{T.}~\bibnamefont{Frederiksen}},
  \bibinfo{author}{\bibfnamefont{T.~N.} \bibnamefont{Todorov}},
  \bibinfo{author}{\bibfnamefont{D.}~\bibnamefont{Dundas}}, \bibnamefont{and}
  \bibinfo{author}{\bibfnamefont{A.~P.} \bibnamefont{Horsfield}},
  \bibinfo{journal}{Physical Review B} \textbf{\bibinfo{volume}{78}},
  \bibinfo{pages}{035446} (\bibinfo{year}{2008}).

\bibitem[{\citenamefont{Ryndyk et~al.}(2009)\citenamefont{Ryndyk, Gutiérrez,
  Song, and Cuniberti}}]{Ryndyk:2008}
\bibinfo{author}{\bibfnamefont{D.~A.} \bibnamefont{Ryndyk}},
  \bibinfo{author}{\bibfnamefont{R.}~\bibnamefont{Gutiérrez}},
  \bibinfo{author}{\bibfnamefont{B.}~\bibnamefont{Song}}, \bibnamefont{and}
  \bibinfo{author}{\bibfnamefont{G.}~\bibnamefont{Cuniberti}}, in
  \emph{\bibinfo{booktitle}{Energy Transfer Dynamics in Biomaterial Systems}},
  edited by \bibinfo{editor}{\bibfnamefont{I.}~\bibnamefont{Burghardt}},
  \bibinfo{editor}{\bibfnamefont{V.}~\bibnamefont{May}},
  \bibinfo{editor}{\bibfnamefont{D.~A.} \bibnamefont{Micha}}, \bibnamefont{and}
  \bibinfo{editor}{\bibfnamefont{E.~R.} \bibnamefont{Bittner}}
  (\bibinfo{publisher}{Springer Berlin Heidelberg}, \bibinfo{year}{2009}),
  vol.~\bibinfo{volume}{93} of \emph{\bibinfo{series}{Springer Series in
  Chemical Physics}}, pp. \bibinfo{pages}{213--335}, ISBN
  \bibinfo{isbn}{978-3-642-02306-4}.

\bibitem[{\citenamefont{Schmidt et~al.}(2008)\citenamefont{Schmidt, Hettler,
  and Sch\"{o}n}}]{Schmidt:2008}
\bibinfo{author}{\bibfnamefont{B.~B.} \bibnamefont{Schmidt}},
  \bibinfo{author}{\bibfnamefont{M.~H.} \bibnamefont{Hettler}},
  \bibnamefont{and}
  \bibinfo{author}{\bibfnamefont{G.}~\bibnamefont{Sch\"{o}n}},
  \bibinfo{journal}{Physical Review B (Condensed Matter and Materials Physics)}
  \textbf{\bibinfo{volume}{77}}, \bibinfo{pages}{165337}
  (\bibinfo{year}{2008}).

\bibitem[{\citenamefont{Tsukada and Mitsutake}(2009)}]{Tsukada:2009}
\bibinfo{author}{\bibfnamefont{M.}~\bibnamefont{Tsukada}} \bibnamefont{and}
  \bibinfo{author}{\bibfnamefont{K.}~\bibnamefont{Mitsutake}},
  \bibinfo{journal}{Journal of the Physical Society of Japan}
  \textbf{\bibinfo{volume}{78}}, \bibinfo{pages}{084701}
  (\bibinfo{year}{2009}).

\bibitem[{\citenamefont{Loos et~al.}(2009)\citenamefont{Loos, Koch, Alvermann,
  Bishop, and Fehske}}]{Loos:2009}
\bibinfo{author}{\bibfnamefont{J.}~\bibnamefont{Loos}},
  \bibinfo{author}{\bibfnamefont{T.}~\bibnamefont{Koch}},
  \bibinfo{author}{\bibfnamefont{A.}~\bibnamefont{Alvermann}},
  \bibinfo{author}{\bibfnamefont{A.~R.} \bibnamefont{Bishop}},
  \bibnamefont{and} \bibinfo{author}{\bibfnamefont{H.}~\bibnamefont{Fehske}},
  \bibinfo{journal}{Journal of Physics: Condensed Matter}
  \textbf{\bibinfo{volume}{21}}, \bibinfo{pages}{395601}
  (\bibinfo{year}{2009}).

\bibitem[{\citenamefont{Avriller and Yeyati}(2009)}]{Avriller:2009}
\bibinfo{author}{\bibfnamefont{R.}~\bibnamefont{Avriller}} \bibnamefont{and}
  \bibinfo{author}{\bibfnamefont{A.~L.} \bibnamefont{Yeyati}},
  \bibinfo{journal}{Physical Review B} \textbf{\bibinfo{volume}{80}},
  \bibinfo{pages}{041309} (\bibinfo{year}{2009}).

\bibitem[{\citenamefont{Haupt et~al.}(2009)\citenamefont{Haupt, Novotn\'y, and
  Belzig}}]{Haupt:2009}
\bibinfo{author}{\bibfnamefont{F.}~\bibnamefont{Haupt}},
  \bibinfo{author}{\bibfnamefont{T.}~\bibnamefont{Novotn\'y}},
  \bibnamefont{and} \bibinfo{author}{\bibfnamefont{W.}~\bibnamefont{Belzig}},
  \bibinfo{journal}{Phys. Rev. Lett.} \textbf{\bibinfo{volume}{103}},
  \bibinfo{pages}{136601} (\bibinfo{year}{2009}).

\bibitem[{\citenamefont{Dash et~al.}(2010)\citenamefont{Dash, Ness, and
  Godby}}]{Dash:2010}
\bibinfo{author}{\bibfnamefont{L.~K.} \bibnamefont{Dash}},
  \bibinfo{author}{\bibfnamefont{H.}~\bibnamefont{Ness}}, \bibnamefont{and}
  \bibinfo{author}{\bibfnamefont{R.~W.} \bibnamefont{Godby}},
  \bibinfo{journal}{Journal of Chemical Physics}
  \textbf{\bibinfo{volume}{132}}, \bibinfo{pages}{104113}
  (\bibinfo{year}{2010}).

\bibitem[{\citenamefont{Ness et~al.}(2010)\citenamefont{Ness, Dash, and
  Godby}}]{Ness:2010}
\bibinfo{author}{\bibfnamefont{H.}~\bibnamefont{Ness}},
  \bibinfo{author}{\bibfnamefont{L.}~\bibnamefont{Dash}}, \bibnamefont{and}
  \bibinfo{author}{\bibfnamefont{R.~W.} \bibnamefont{Godby}},
  \bibinfo{journal}{Physical Review B} \textbf{\bibinfo{volume}{82}},
  \bibinfo{pages}{085426} (\bibinfo{year}{2010}).

\bibitem[{\citenamefont{Wang and Thoss}(2011)}]{Wang_H:2011}
\bibinfo{author}{\bibfnamefont{H.}~\bibnamefont{Wang}} \bibnamefont{and}
  \bibinfo{author}{\bibfnamefont{M.}~\bibnamefont{Thoss}},
  \emph{\bibinfo{title}{Numerically exact, time-dependent treatment of
  vibrationally coupled electron transport in single-molecule junctions}}
  (\bibinfo{year}{2011}), \eprint[arXiv]{arXiv:1103.4945v1}.

\bibitem[{\citenamefont{Garcia-Lekue et~al.}(2011)\citenamefont{Garcia-Lekue,
  Sanchez-Portal, Arnau, and Frederiksen}}]{Garcia-Lekue:2011}
\bibinfo{author}{\bibfnamefont{A.}~\bibnamefont{Garcia-Lekue}},
  \bibinfo{author}{\bibfnamefont{D.}~\bibnamefont{Sanchez-Portal}},
  \bibinfo{author}{\bibfnamefont{A.}~\bibnamefont{Arnau}}, \bibnamefont{and}
  \bibinfo{author}{\bibfnamefont{T.}~\bibnamefont{Frederiksen}},
  \emph{\bibinfo{title}{Simulation of inelastic electron tunneling spectroscopy
  of single molecules with functionalized tips}} (\bibinfo{year}{2011}),
  \eprint[arXiv]{arXiv:1103.4302v1}.

\bibitem[{\citenamefont{Ueda et~al.}(2011)\citenamefont{Ueda, Entin-Wohlman,
  and Aharony}}]{Ueda:2011}
\bibinfo{author}{\bibfnamefont{A.}~\bibnamefont{Ueda}},
  \bibinfo{author}{\bibfnamefont{O.}~\bibnamefont{Entin-Wohlman}},
  \bibnamefont{and} \bibinfo{author}{\bibfnamefont{A.}~\bibnamefont{Aharony}},
  \emph{\bibinfo{title}{Effects of coupling to vibrational modes on the ac
  conductance of molecular junctions}} (\bibinfo{year}{2011}),
  \eprint[arXiv]{arXiv:1101.4440v1}.

\bibitem[{\citenamefont{Meir and Wingreen}(1992)}]{Meir:1992}
\bibinfo{author}{\bibfnamefont{Y.}~\bibnamefont{Meir}} \bibnamefont{and}
  \bibinfo{author}{\bibfnamefont{N.~S.} \bibnamefont{Wingreen}},
  \bibinfo{journal}{Physical Review Letters} \textbf{\bibinfo{volume}{68}},
  \bibinfo{pages}{2512} (\bibinfo{year}{1992}).

\bibitem[{\citenamefont{Datta et~al.}(1997)\citenamefont{Datta, Tian, Hong,
  Reifenberger, Henderson, and Kubiak}}]{Datta:1997}
\bibinfo{author}{\bibfnamefont{S.}~\bibnamefont{Datta}},
  \bibinfo{author}{\bibfnamefont{W.~D.} \bibnamefont{Tian}},
  \bibinfo{author}{\bibfnamefont{S.~H.} \bibnamefont{Hong}},
  \bibinfo{author}{\bibfnamefont{R.}~\bibnamefont{Reifenberger}},
  \bibinfo{author}{\bibfnamefont{J.~I.} \bibnamefont{Henderson}},
  \bibnamefont{and} \bibinfo{author}{\bibfnamefont{C.~P.}
  \bibnamefont{Kubiak}}, \bibinfo{journal}{Physical Review Letters}
  \textbf{\bibinfo{volume}{79}}, \bibinfo{pages}{2530} (\bibinfo{year}{1997}).

\end{thebibliography}

\end{document}